\documentclass[letterpaper]{article} 
\usepackage{aaai24}  
\usepackage{times}  
\usepackage{helvet}  
\usepackage{courier}  
\usepackage[hyphens]{url}  
\usepackage{graphicx} 
\usepackage{natbib}  
\usepackage{caption} 
\usepackage{algorithm}
\usepackage[noend]{algorithmic}

\usepackage{newfloat}
\usepackage{listings}
\DeclareCaptionStyle{ruled}{labelfont=normalfont,labelsep=colon,strut=off} 
\floatstyle{ruled}
\newfloat{listing}{tb}{lst}{}
\floatname{listing}{Listing}
\title{Learning Coalition Structures with Games}
\author{
    Yixuan Even Xu,\textsuperscript{\rm 1}\thanks{This work was done when Xu was a visiting intern at CMU.}\quad 
    Chun Kai Ling,\textsuperscript{\rm 2,3}\quad 
    Fei Fang\textsuperscript{\rm 3}
}
\affiliations{
    \textsuperscript{\rm 1}Tsinghua University,\quad 
    \textsuperscript{\rm 2}Columbia University ,\quad 
    \textsuperscript{\rm 3}Carnegie Mellon University\\

    xuyx20@mails.tsinghua.edu.cn,\quad 
    \{chunkail,feif\}@cs.cmu.edu
}

\usepackage{enumerate}

\usepackage{subcaption}

\usepackage{amsthm}
\usepackage{amsmath}
\usepackage{amssymb}
\usepackage{mathtools}

\usepackage[capitalise]{cleveref}

\usepackage{threeparttable}
\usepackage{hhline}
\usepackage{multirow}

\usepackage{tikz}
\usetikzlibrary{calc}

\usepackage{thmtools}
\usepackage{thm-restate}
\newtheorem{theorem}{Theorem}[section]
\newtheorem{definition}{Definition}[section]

\newtheorem{corollary}{Corollary}[section]

\newtheorem{proposition}{Proposition}[section]

\newtheorem{lemma}{Lemma}[section]

\newenvironment{proofof}[1]{{\bf \noindent Proof of #1:  }}{\hfill\rule{2mm}{2mm}}

\newcommand{\N}{N}
\newcommand{\A}{\mathcal A}
\renewcommand{\v}{\mathbf v}
\renewcommand{\r}{\mathbf r}
\newcommand{\x}{\mathbf x}
\newcommand{\y}{\mathbf y}
\newcommand{\V}{\mathcal V}
\newcommand{\U}{\mathcal U}
\renewcommand{\S}{\mathcal S}
\newcommand{\M}{\mathcal M}
\newcommand{\G}{\mathcal G}
\renewcommand{\O}{\mathcal O}
\newcommand{\NF}{\mathcal{N}}
\newcommand{\ind}[1]{\mathbb{I}\hspace{-0.1em}\left[\vphantom{\sum}#1\right]}

\begin{document}

\maketitle

\begin{abstract}
    Coalitions naturally exist in many real-world systems involving multiple decision makers such as ridesharing, security, and online ad auctions, but the coalition structure among the agents is often unknown. We propose and study an important yet previously overseen problem -- Coalition Structure Learning (CSL), where we aim to carefully design a series of games for the agents and infer the underlying coalition structure by observing their interactions in those games. We establish a lower bound on the sample complexity  -- defined as the number of games needed to learn the structure -- of any algorithms for CSL and propose the Iterative Grouping (IG) algorithm for designing normal-form games to achieve the lower bound. We show that IG can be extended to other succinct games such as congestion games and graphical games. Moreover, we solve CSL in a more restrictive and practical setting: auctions. We show a variant of IG to solve CSL in the auction setting even if we cannot design the bidder valuations. Finally, we conduct experiments to evaluate IG in the auction setting and the results align with our theoretical analysis.
\end{abstract}

\section{Introduction}
\label{sec:introduction}
Coalitions are an integral part of large, multi-agent environments. 
Some coalitions can lead to undesirable outcomes. 
For example, in ridesharing platforms (e.g., Uber, Lyft), groups of drivers sometimes deliberately and simultaneously disconnect themselves from the platform in hopes of artificially inducing a price surge which they enjoy later at the expense of the platform and riders \cite{hamilton2019uber,Sweeney2019manipulate,Dowling2023drive}, sparking studies on mechanisms to discourage such behaviors \cite{Tripathy2022}. In security domains, coordinated attacks are often more difficult to mitigate compared to those conducted in isolation. \cite{jena2021design,lakshminarayana2019moving}. 
On the other hand, coalitions are common and crucial to the proper functioning of real-world societies.


Ultimately, knowing the underlying coalition structure in such environments can lead to more accurate game models, more robust strategies, or the construction of better welfare-maximizing mechanisms.
However, unlike payoffs, it is often not known apriori which coalitions (if any) exist. As such, we propose the \textit{Coalition Structure Learning} (CSL) problem, where we actively put agents through a small set of carefully designed games and infer the underlying coalition structure by observing their behavior.

We stress the difference between our work and cooperative game theory. Our work identifies coalition structures by exploiting the differences in interactions between agents and is separate from the study of underlying mechanisms ensuring the stability of the said coalitions.

In this paper, we assume members in a coalition secretly share their individual utilities, i.e., they act as a joint agent whose utility equals the sum of the individual utilities of its members. 
Crucially, this difference in behavior allows us to detect coalitions.
Consider the game shown in \cref{subfig:coalition_a}, a variant of the classic Prisoner's Dilemma. Here, the only Nash Equilibrium (NE) is for both agents to \textbf{D}efect. However, if they are in a coalition, they behave collectively as a single agent with payoffs shown in \cref{subfig:coalition_b}. From the coalition's perspective, it is rational for both agents to \textbf{C}ooperate as it maximizes the sum of both agent's payoff.

 \begin{figure}[htbp]
     \centering
     \begin{subfigure}[b]{0.18\textwidth}
         \begin{equation*}
             \begin{array}{|c|c|c|}
                 \hline & \textup{C}_y & \textup{D}_y \\
                 \hline \textup{C}_x & (3,3) & (0,5) \\
                 \hline \textup{D}_x & (5,0) & \mathbf{(1,1)} \\
                 \hline
             \end{array}
         \end{equation*}
         \caption{Not in a coalition}
         \label{subfig:coalition_a}
     \end{subfigure}
     \hfill
     \begin{subfigure}[b]{0.27\textwidth}
         \begin{equation*}
             \begin{array}{|c|c|c|c|}
                 \hline \textup{C}_x\textup{C}_y & \textup{C}_x\textup{D}_y & \textup{D}_x\textup{C}_y & \textup{D}_x\textup{D}_y\\
                 \hline \mathbf{3+3} & 0+5 & 5+0 & 1+1 \\
                 \hline
             \end{array}
         \end{equation*}
         \caption{In a coalition}
         \label{subfig:coalition_b}
     \end{subfigure}
     \caption{A variant of Prisoner's dilemma when agents $x$ and $y$ are (b) in and (a) not in a coalition. 
     Bolded cells are the (unique) Nash Equilibria. 
     }
     \label{fig:coalition}
\end{figure}

More generally, we have a set $\N = \{1,2,\dots,n\}$ of $n$ strategic agents\footnote{We provide a list of key notations in \cref{app:list_of_notations}.}, divided into $m$ separate coalitions. A coalition $S\subseteq \N$ is a nonempty subset of the agents, in which the agents
coordinate with each other. 
A coalition structure of the agents is represented by a partition $\S=\{S_1,S_2,\dots,S_m\}$ of $\N$, where $S_1,S_2,\dots,S_m$ are mutually disjoint coalitions and $\bigcup_{i=1}^m S_i = \N$. Note that some of the coalitions might be singletons. We use $[i]_\S$ to denote the coalition that agent $i$ belongs to under $\S$.  If $[i]_\S = \{i\}$ for each $i \in \N$, we recover the regular game setting.

In CSL, both $m$ and $\S$ are unknown, and the goal is to recover them by observing how the agents interact with each other in a series of designed games. At each timestep, we present a game $\G$ to the agents and make an observation $\O$ about the equilibria in $\G$. 
As shown in \cref{fig:coalition}, different coalition structures will lead to different sets of equilibria, which makes CSL possible to solve.
We restrict $\O$ to be a single-bit oracle, indicating whether a pre-specified strategy profile $\Sigma$ is a Nash Equilibrium of $\G$. 
This is the simplest observation to make and can be implemented in practice by presenting $\Sigma$ as a default strategy profile to the agents and observing whether \textit{any} agent deviates from it. 

We define the \textit{sample complexity} of an algorithm on a CSL instance as the number of games it presents to agents before the correct coalition structure is learned. We are interested in algorithms with low sample complexity.

In this paper, we thoroughly study CSL with the single-bit observation oracle $\O$. In many real-world settings, there will be restrictions on what kind of games can be designed and presented to the agents. Therefore, we study CSL under various settings of the class of games that $\G$ belongs to. Specifically, we make the following contributions: \textbf{(1).} We propose and formally model the CSL problem. \textbf{(2).}  
We show a lower bound of sample complexity as a function of the number of agents $n$ for algorithms solving CSL (\cref{thm:lower_bound_of_sample_complexity}). 
\textbf{(3).} We propose our Iterative Grouping (IG) algorithm for solving CSL when $\G$ is restricted to normal form games (\cref{alg:normal_form_games}) and show that it achieves the optimal sample complexity up to low order terms (\cref{thm:normal_form_games}). \textbf{(4).} We extend IG to solve CSL with congestion games and graphical games, again with optimal sample complexity (\cref{subsec:extension_to_other_succinct_games}). \textbf{(5).} We propose AuctionCSL, a variant of CSL in the grounded setting of second-price auctions with personalized reserve prices, and extend IG to solve AuctionCSL (\cref{sec:csl_with_auctions}). \textbf{(6).} We extensively conduct experiments to evaluate IG in the auction setting (\cref{sec:experiments}). The experiments align with our theoretical results, showing that IG is a practical approach to AuctionCSL. Below we summarize the theoretical results of this paper in \cref{table:summary}.

\begin{table}[htbp]
	\centering
	\begin{tabular}{|c|c|c|c|c|}
		\hline
		Setting & Sample Complexity & Section \\
		\hline
        Lower Bound & $(1 - o(1))n \log_2 n$ & \cref{subsec:lower_bound_of_sample_complexity} \\
		\hline
        Normal Form & $n \log_2 n+3n$ & \cref{subsec:learning_algorithm} \\
        \hline
        Congestion & $n \log_2 n+3n$ & \cref{subsec:extension_to_other_succinct_games} \\
		\hline
        Graphical & $n \log_2 n+3n$ & \cref{app:csl_with_graphical_games} \\
        \hline
        Auction & $(4.16+o(1))n \log_2 n$ & \cref{sec:csl_with_auctions} \\
        \hline
	\end{tabular}
	\caption{Summary of theoretical results.}
	\label{table:summary}
\end{table}

\section{Related Work}
\label{sec:related_work}
In recent years, there has been significant interest in the learning of games. One such direction is \textit{Inverse Game Theory}, which seeks to compute game parameters (e.g., agent utilities, chance) that give rise to a particular empirically observed equilibrium \cite{DBLP:conf/icml/WaughZB11,kuleshov2015inverse,DBLP:conf/ijcai/LingFK18,geiger2021learning,Peng_Shen_Tang_Zuo_2019,letchford2009learning}.
In an ``active'' setting closer to our work, \citet{balcan2015commitment,DBLP:conf/ijcai/HaghtalabFNSPT16} show that attacker utilities in Stackelberg security games may be learned by observing best-responses to chosen defender strategies. 
More broadly, the field of \textit{Empirical Game-Theoretic Analysis} reasons about games and their structure by interleaving game simulation and analysis \cite{wellman2006methods}. Another related direction is given by \citet{athey2002identification}, who identify different auctions based on winning bids or bidders. Recent work by \citet{kenton2023discovering} distinguishes between agents and the environment by extending techniques from causal inference.
In all of these works, the focus is to learn agent payoffs and other game parameters (e.g., chance probabilities, item valuations, and distributions), assuming that agents and any coalitions are pre-specified. In contrast, CSL learns \textit{coalition structures} given the freedom to design agent payoffs or other game parameters.
Finally, \citet{mazrooei2013automating} and \citet{bonjour2022information} detect the existence of a single coalition, but not the entire coalition structure in multiplayer games.

\section{CSL with Normal Form Games}
\label{sec:csl_with_normal_form_games}

In this section, we present how to solve the CSL problem when $\mathcal{G}$ is restricted to the set of all normal form games. We assume in this section that we have the power to design the whole game matrix. This section demonstrates the main idea of the paper, which will be recurring in more complicated and restricted settings in \cref{sec:csl_with_auctions}.

\subsection{Lower Bound of Sample Complexity}
\label{subsec:lower_bound_of_sample_complexity}

We start our investigation with a lower bound of the sample complexity of any algorithm that solves the CSL problem. It serves as a reference for designing future algorithms.

\begin{theorem}
    \label{thm:lower_bound_of_sample_complexity}
    An algorithm solving the CSL problem has a sample complexity of at least $n \log_2 n - O(n\log_2\log_2 n)$.
\end{theorem}

\begin{proofof}{\cref{thm:lower_bound_of_sample_complexity}}
    For every game $\G$ presented to the agents, we get at most 1 bit of information from $\O$. The number of possible partitions of $\N$ is the Bell number $B_n$. Therefore, to distinguish between all possible partitions, we need at least $\left\lceil \log_2 B_n\right\rceil=n \log_2 n - O(n\log_2\log_2 n)$ bits of information, which follows from the asymptotic expression of Bell number established in \citet{de1981asymptotic}. 
\end{proofof}

\subsection{Pairwise Testing via Normal-Form Gadgets}
\label{subsec:gadget_construction_normal_form}

Let $\S^*$ be the ground truth coalition structure. It is useful to consider the problem of determining if a given pair of agents $(x, y)$ are in the same coalition, i.e., $[x]_\mathcal{S^*}=[y]_\mathcal{S^*}$. The solution to this subproblem is given by a \textit{normal-form gadget} game inspired by \cref{fig:coalition}, and forms the building block toward our eventual Iterative Grouping algorithm.


\begin{definition}
    \label{def:game-strategy-pair}
    A game-strategy pair $(\G, \Sigma)$ is a $n$-player normal form game with $\Sigma$ as a \textbf{default strategy profile} in $\G$. 
\end{definition}

\begin{definition}
    \label{def:normal_form_gadget}
    A \textbf{normal form gadget} $\NF(x,y)=(\G, \Sigma)$ 
    is a game-strategy pair where players $i\in\N\setminus\{x,y\}$ are dummies with one action $\textup{D}_i$ and recieve $0$ utility. Players $x$ and $y$ have actions $\{\textup{C}_x,\textup{D}_x\}$ and $\{\textup{C}_y,\textup{D}_y\}$ and utilities
    shown in \cref{subfig:coalition_a}.
    The \textbf{default strategy profile} is $\Sigma = (\textup{D}_1,\dots,\textup{D}_n)$.
\end{definition}


\begin{lemma}
    \label{lem:normal_form_gadget}
    The default strategy profile of $\NF(x, y)$ 
    is a Nash Equilibrium if and only if $[x]_\mathcal{S^*} \neq [y]_\mathcal{S^*}$. 
\end{lemma}

\begin{proofof}{\cref{lem:normal_form_gadget}}
    If $x$ and $y$ are in the same coalition, they act as a joint player with utility equal to the sum of their individual utilities. Then, by deviating to $(\textup{C}_x,\textup{C}_y)$, the utility of the joint player will increase from $2$ to $6$. Thus the default strategy profile is not a Nash Equilibrium. If $x$ and $y$ are in different coalitions, then the unilateral deviation of either the coalition of $x$ or $y$ will not increase their utility. Thus the default strategy profile is a Nash Equilibrium.
\end{proofof}

We remark that the game in \cref{subfig:coalition_a} is not the only game that can be used to construct the normal form gadget in \cref{def:normal_form_gadget}. A game with a unique NE from which agents have incentives to deviate when they are in the same coalition would serve the purpose.
\cref{def:normal_form_gadget} and \cref{lem:normal_form_gadget} shows how to detect pairwise coalition. With \cref{lem:normal_form_gadget}, we can already solve CSL with a sample complexity of $\frac{1}{2}n(n-1)$ by querying the observation oracle $\O(\NF(x,y))$ for all $1\leq x < y \leq n$. However, we can do better by checking multiple agent pairs at the same time, as detailed next. 

\subsection{The Iterative Grouping Algorithm}
\label{subsec:learning_algorithm}

\begin{figure*}[htbp]
    \centering
    
    \begin{subfigure}[b]{0.32\textwidth}
        \centering
        \begin{tikzpicture}[scale=1]
            \node[circle, draw] (1) at (0,1.2) {1};
            \node[circle, draw] (2) at (1.5,1.2) {2};
            \node[circle, draw] (3) at (3,1.2) {3};
            \node[circle, draw] (4) at (0,0) {4};
            \node[circle, draw] (5) at (1.5,0) {5};
            \node[circle, draw] (6) at (3,0) {6};
            
            \draw[rounded corners, dashed] ($(1.north west)+(-0.15,0.15)$) rectangle ($(1.south east)+(0.15,-0.15)$);
            \draw[rounded corners, dashed] ($(2.north west)+(-0.15,0.15)$) rectangle ($(3.south east)+(0.15,-0.15)$);
            \draw[rounded corners, dashed] ($(4.north west)+(-0.15,0.15)$) rectangle ($(6.south east)+(0.15,-0.15)$);
            \draw[rounded corners, dashed] ($(3.north east)+(0.3,0.3)$) -- ($(2.north west)+(-0.3,0.3)$) -- ($(5.north west)+(-0.3,0.3)$) -- ($(4.north west)+(-0.3,0.3)$) -- ($(4.south west)+(-0.3,-0.3)$) -- ($(6.south east)+(0.3,-0.3)$) -- cycle;

            \node at (0.55,1.35) {$i$};
            \node at (2.25,1.2) {$T_\alpha$};
            \node at (2.25,0) {$T_\beta$};
            \node at (2.25,0.6) {$T$};

            \draw (1) -- (2);
            \draw (2) -- (4);
            \draw (1) -- (4);
            \draw (3) -- (6);
        \end{tikzpicture}
		\label{subfigure:binary_a}
    \end{subfigure}
    \begin{subfigure}[b]{0.32\textwidth}
        \centering
        \begin{tikzpicture}[scale=1]
            \node[circle, draw] (1) at (0,1.2) {1};
            \node[circle, draw] (2) at (1.5,1.2) {2};
            \node[circle, draw] (3) at (3,1.2) {3};
            \node[circle, draw] (4) at (0,0) {4};
            \node[circle, draw] (5) at (1.5,0) {5};
            \node[circle, draw] (6) at (3,0) {6};
            
            \draw[rounded corners, dashed] ($(1.north west)+(-0.15,0.15)$) rectangle ($(1.south east)+(0.15,-0.15)$);
            \draw[rounded corners, dashed] ($(2.north west)+(-0.15,0.15)$) rectangle ($(2.south east)+(0.15,-0.15)$);
            \draw[rounded corners, dashed] ($(3.north west)+(-0.15,0.15)$) rectangle ($(3.south east)+(0.15,-0.15)$);
            \draw[rounded corners, dashed] ($(2.north west)+(-0.3,0.3)$) rectangle ($(3.south east)+(0.3,-0.3)$);

            \node at (0.55,1.35) {$i$};
            \node at (2.25,1.2) {$T$};
            \node at (1.5,0.6) {$T_\alpha$};
            \node at (3,0.6) {$T_\beta$};

            \draw (1) -- (2);
            \draw (2) -- (4);
            \draw (1) -- (4);
            \draw (3) -- (6);
        \end{tikzpicture}
		\label{subfigure:binary_b}
    \end{subfigure}
    \begin{subfigure}[b]{0.32\textwidth}
        \centering
        \begin{tikzpicture}[scale=1]
            \node[circle, draw] (1) at (0,1.2) {1};
            \node[circle, draw] (2) at (1.5,1.2) {2};
            \node[circle, draw] (3) at (3,1.2) {3};
            \node[circle, draw] (4) at (0,0) {4};
            \node[circle, draw] (5) at (1.5,0) {5};
            \node[circle, draw] (6) at (3,0) {6};
            
            \draw[rounded corners, dashed] ($(1.north west)+(-0.15,0.15)$) rectangle ($(1.south east)+(0.15,-0.15)$);
            \draw[rounded corners, dashed] ($(2.north west)+(-0.15,0.15)$) rectangle ($(2.south east)+(0.15,-0.15)$);

            \node at (0.55,1.35) {$i$};
            \node at (2.05,1.2) {$T$};

            \draw (1) -- (2);
            \draw (2) -- (4);
            \draw (1) -- (4);
            \draw (3) -- (6);
        \end{tikzpicture}
		\label{subfigure:binary_c}
    \end{subfigure}
    
    \caption{Example of the binary search process in \cref{alg:normal_form_games} (Lines 4 to 11). The ground truth coalition structure is $\S^* = \{\{1, 2, 4\}, \{3, 6\}, \{5\}\}$ as shown by the solid lines. The algorithm is trying to find an agent in agent $1$'s coalition. At first $T = \{2,3,4,5,6\}$ and is partitioned into $T_\alpha = \{2,3\}$ and $T_\beta = \{4,5,6\}$ (left). As $T_\alpha$ contains an agent in $1$'s coalition, $T$ is replaced by $T_\alpha$ and then partitioned into $T_\alpha = \{2\}$ and $T_\beta = \{3\}$ (middle). Then, as $T_\alpha$ still contains an agent in $1$'s coalition, $T$ is replaced by $T_\alpha = \{2\}$ and we find an agent $2$ in agent $1$'s coalition (right).}
	\label{fig:binary}
\end{figure*}
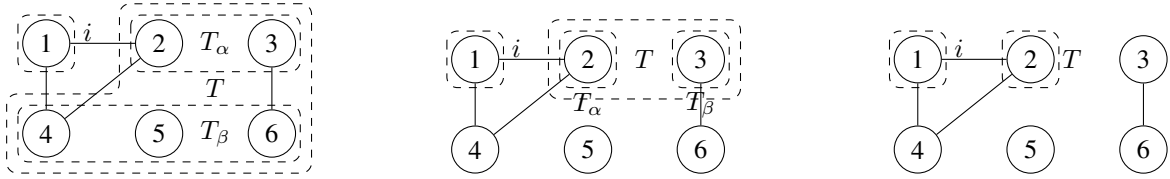

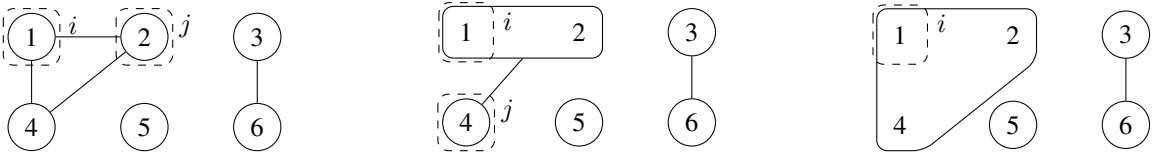
\begin{figure*}[htbp]
    \centering
    
    \begin{subfigure}[b]{0.32\textwidth}
        \centering
        \begin{tikzpicture}[scale=1]
            \node[circle, draw] (1) at (0,1.2) {1};
            \node[circle, draw] (2) at (1.5,1.2) {2};
            \node[circle, draw] (3) at (3,1.2) {3};
            \node[circle, draw] (4) at (0,0) {4};
            \node[circle, draw] (5) at (1.5,0) {5};
            \node[circle, draw] (6) at (3,0) {6};
            
            \draw[rounded corners, dashed] ($(1.north west)+(-0.15,0.15)$) rectangle ($(1.south east)+(0.15,-0.15)$);
            \draw[rounded corners, dashed] ($(2.north west)+(-0.15,0.15)$) rectangle ($(2.south east)+(0.15,-0.15)$);

            \node at (0.55,1.35) {$i$};
            \node at (2.05,1.35) {$j$};

            \draw (1) -- (2);
            \draw (2) -- (4);
            \draw (1) -- (4);
            \draw (3) -- (6);
        \end{tikzpicture}
		\label{subfigure:merger_a}
    \end{subfigure}
    \begin{subfigure}[b]{0.32\textwidth}
        \centering
        \begin{tikzpicture}[scale=1]
            \node (1) at (0,1.2) {1};
            \node (2) at (1.5,1.2) {2};
            \node[circle, draw] (3) at (3,1.2) {3};
            \node[circle, draw] (4) at (0,0) {4};
            \node[circle, draw] (5) at (1.5,0) {5};
            \node[circle, draw] (6) at (3,0) {6};
            
            \draw[rounded corners] ($(1.north west)+(-0.10,0.10)$) rectangle ($(2.south east)+(0.10,-0.10)$);
            \draw[rounded corners, dashed] ($(1.north west)+(-0.15,0.15)$) rectangle ($(1.south east)+(0.15,-0.15)$);
            \draw[rounded corners, dashed] ($(4.north west)+(-0.15,0.15)$) rectangle ($(4.south east)+(0.15,-0.15)$);

            \node at (0.55,1.35) {$i$};
            \node at (0.55,0.15) {$j$};

            \draw (0.75, 0.85) -- (4);
            \draw (3) -- (6);
        \end{tikzpicture}
		\label{subfigure:merger_b}
    \end{subfigure}
    \begin{subfigure}[b]{0.32\textwidth}
        \centering
        \begin{tikzpicture}[scale=1]
            \node (1) at (0,1.2) {1};
            \node (2) at (1.5,1.2) {2};
            \node[circle, draw] (3) at (3,1.2) {3};
            \node (4) at (0,0) {4};
            \node[circle, draw] (5) at (1.5,0) {5};
            \node[circle, draw] (6) at (3,0) {6};
            
            \draw[rounded corners, dashed] ($(1.north west)+(-0.15,0.15)$) rectangle ($(1.south east)+(0.15,-0.15)$);
            \draw[rounded corners] ($(1.north west)+(-0.10,0.10)$) -- ($(2.north east)+(0.10,0.10)$) -- ($(2.south east)+(0.10,-0.10)$) -- ($(4.south east)+(0.10,-0.10)$) -- ($(4.south west)+(-0.10,-0.10)$) --  cycle;

            \node at (0.55,1.35) {$i$};

            \draw (3) -- (6);
        \end{tikzpicture}
		\label{subfigure:merger_c}
    \end{subfigure}
    \caption{Example of one iteration of the outer for loop in \cref{alg:normal_form_games} (Lines 3 to 12). The ground truth coalition structure is $\S^* = \{\{1, 2, 4\}, \{3, 6\}, \{5\}\}$ as shown by the solid lines. The algorithm is trying to find all agents in agent $1$'s coalition. After finding agent $2$ in agent $1$'s coalition, the algorithm merges their coalitions $\{1\},\{2\}$ into one coalition $\{1,2\}$ (left). Then, the algorithm finds agent $4$ in agents $1$ and $2$'s coalition and merges their coalitions $\{1,2\}$ and $\{4\}$ into one coalition $\{1,2,4\}$ (middle). Finally, the algorithm confirms that agent $1$'s coalition is finalized (right).}
	\label{fig:merger}
\end{figure*}

Our Iterative Grouping (IG) algorithm solves CSL with a sample complexity matching the bound in \cref{thm:lower_bound_of_sample_complexity}.
IG begins with an initial coalition structure where each agent is in a separate coalition. Then, for agent $i$, IG iteratively tries to find another agent $j$ within $i$'s coalition. If it finds such an agent, it merges $i$ and $j$'s coalitions. Otherwise, it finalizes $i$'s coalition and moves on to the next agent. In either case, the number of unfinalized coalitions decreases. Therefore, IG will eventually find the correct coalition structure. 

To find such an agent $j$, IG uses a method similar to binary search. Specifically, we will introduce in \cref{lem:product_of_normal_form_game_gadget} a way that allows us to use the observation of a \textit{single} game to determine for a set $T \subseteq \N$, whether there is an agent $j\in T$ that is also within $i$'s coalition, i.e., whether $T \cap [i]_{\S^*}\ne \varnothing$. If so, we bisect $T$ into two sets $T_\alpha$ and $T_\beta$ and use another game to determine which of the two sets $j$ is in. We then repeat this process recursively to locate $j$ efficiently. 

With that in mind, we proceed to describe IG formally. We start by defining the product of game-strategy pairs, which returns a game equivalent to \textit{playing the two games separately with utilities of each player summed}, as well as a product of default strategy profiles for each player.

\begin{definition}
    \label{def:product_of_mixed_strategies}
    Let $\sigma_1 = (c_1,\dots,c_{k_1}),\sigma_2 = (d_1,\dots,d_{k_2})$ be two mixed strategies over the sets of actions $A = \{a_1,\dots,a_{k_1}\},B = \{b_1,\dots,b_{k_2}\}$ respectively, where $c_{\theta},d_{\eta}$ are the probabilities of choosing $a_{\theta},b_{\eta}$ respectively. The \textbf{product} of $\sigma_1$ and $\sigma_2$ is a mixed strategy $\sigma_1\times\sigma_2$ over $A\times B$, where the probability of choosing $(a_{\theta},b_{\eta})$ is $c_{\theta}d_{\eta}$.
\end{definition}

\begin{definition}
    \label{def:product_of_normal_form_games}
    Let $(\G_1,\Sigma_1),(\G_2, \Sigma_2)$ be two game-strategy pairs where $A_{x,i},u_{x,i}$ are the action sets and utility function of player $i$ in $\G_x$ respectively for $x\in\{1,2\}$. Let $\Sigma_1 = (\sigma_{1,i})_{i\in\N}$ and $\Sigma_2 = (\sigma_{2,i})_{i\in\N}$. 
    The \textbf{product} of $(\G_1, \Sigma_1)$ and $(\G_2, \Sigma_2)$ is a game-strategy pair $(\G_p, \Sigma_p)$. Here, $\G_p$ is a normal form game with action set $A_{1,i}\times A_{2,i}$ and utility function $u_{1,i} + u_{2,i}$ for each player $i\in N$. $\Sigma_p = (\sigma_{1,i}\times\sigma_{2,i})_{i\in\N}$.
\end{definition}

Then, by querying the observation oracle for the product of several games, we will get an aggregated observation. We formalize this idea in the following lemma.

\begin{lemma}
    \label{lem:product_of_normal_form_game_gadget}
    Let $\{\NF(x_\theta, y_\theta) = (\G_\theta, \Sigma_\theta)\mid \theta\in\{1,\dots,k\}\}$ be a set of $k$ normal form gadgets. The default strategy profile of $\NF(x_1,y_1)\times\cdots\times\NF(x_k,y_k)$ is a Nash Equilibrium if and only if for each $\theta\in\{1,2,\dots,k\}$, $[x_\theta]_{\S^*}\ne [y_\theta]_{\S^*}$.
\end{lemma}

\begin{proofof}{\cref{lem:product_of_normal_form_game_gadget}}
    By \cref{def:product_of_normal_form_games}, playing the product game is equivalent to separately playing $\G_1,\dots,\G_k$, and sum up the resulting utilities of each player. Therefore, the default strategy profile of the product is a Nash Equilibrium if and only if the default strategy profile of each $\G_i$ is a Nash Equilibrium. Applying \cref{lem:normal_form_gadget} completes the proof. 
\end{proofof}

With \cref{lem:product_of_normal_form_game_gadget}, we can design a more efficient Iterative Grouping algorithm for the CSL problem (\cref{alg:normal_form_games}).

\begin{algorithm}[htbp]
    \caption{Iterative Grouping (IG)}
    \label{alg:normal_form_games}

    \textbf{Input}: The number of agents $n$ and an observation oracle $\O$

    \textbf{Output}: A coalition structure $\S$ of the agents

    \begin{algorithmic}[1] 
        \STATE Let $\S\gets\{\{1\},\{2\},\dots,\{n\}\}$.
        \FOR {$i\in\N$}
            \WHILE {$\O(\prod_{[j]_\S \ne [i]_\S}\NF(i,j)) = \textup{false}$}
                \STATE Let $T\gets\{j\in\N\mid [j]_\S \ne [i]_\S\}$.
                \WHILE {$|T| > 1$}
                    \STATE Partition $T$ into $T_\alpha,T_\beta$ where $\left||T_\alpha|-|T_\beta|\right|\leq 1$.
                    \IF {$\O(\prod_{j\in T_\alpha}\NF(i,j)) = \textup{false}$}
                        \STATE Let $T\gets T_\alpha$.
                    \ELSE
                        \STATE Let $T\gets T_\beta$.
                    \ENDIF
                \ENDWHILE
                \STATE Let $j\gets \textup{the only element in $T$}$.
                \STATE Merge $[i]_\S$ and $[j]_\S$ in $\S$.
            \ENDWHILE
        \ENDFOR
        \STATE \textbf{return} $\S$.
    \end{algorithmic}
\end{algorithm}

IG (\cref{alg:normal_form_games}) starts with the initial coalition structure $\S=\{\{1\},\{2\},\dots,\{n\}\}$, where each agent is in a separate coalition (Line 1). In each iteration of the outer for loop (Lines 3 to 12), we consider an agent $i$ and try to find all agents in $i$'s coalition $[i]_{\S^*}$, where $\S^*$ is the ground truth coalition structure. 
In Line 3, we present a game $\prod_{[j]_\S \ne [i]_\S}\NF(i,j)$ to the agents, where we concurrently ask each agent $j$ that is not currently recognized as in $i$'s coalition $[i]_\S$ to play the normal form gadget $\NF(i,j)$ with $i$.
If the default strategy profile in this game is not a Nash Equilibrium (Line 3), then according to \cref{lem:product_of_normal_form_game_gadget}, there must be an agent outside of $[i]_\S$ that is in the same coalition with $i$. We use binary search (Lines 4 to 10) to locate this agent $j$ (\cref{fig:binary}) and merge $i$ and $j$'s coalitions (Lines 11 to 12). This is repeated until all players in $i$'s coalition are found (\cref{fig:merger}). Repeating this for all players $i \in \N$ guarantees we get $\S = \S^*$ once IG terminates. 

\begin{restatable}{theorem}{normalformgames}
    \label{thm:normal_form_games}
    IG solves the CSL problem with a sample complexity upper bounded by $n \log_2 n + 3n$.
\end{restatable}

The proof of \cref{thm:normal_form_games} is deferred to \cref{appsub:proof_of_thm_normal_form_games}. Combined with \cref{thm:lower_bound_of_sample_complexity}, \cref{thm:normal_form_games} shows that IG solves the CSL problem with optimal sample complexity and a matching constant up to low order terms.

\subsection{Extension to Other Succinct Games}
\label{subsec:extension_to_other_succinct_games}

IG solves CSL with normal form games. However, sometimes there are external restrictions on what kind of games we can design and present to the agents, forbidding us from using general normal form games.
Thus, in this subsection, we briefly discuss how to extend IG to other succinct games, like congestion games and graphical games.

\paragraph{CSL with congestion games.} IG can also be extended to congestion games \cite{rosenthal1973class} with a modified gadget construction. For a pair of players $x$ and $y$, we define the congestion game gadget as the congestion game below.
\begin{center}
    \begin{tikzpicture}[scale=1]
        \node[circle, draw] (S) at (0,0) {$S$};
        \node[circle, draw] (1) at (1.5,0.5) {1};
        \node[circle, draw] (2) at (1.5,-0.5) {2};
        \node[circle, draw] (T) at (3,0) {$T$};
    
        \node at (0.6,0.45) {$c$};
        \node at (0.6,-0.45) {$2.5$};
        \node at (2.4,0.45) {$2.5$};
        \node at (2.4,-0.45) {$c$};

        \draw[->, >=latex, line width=0.5pt] (1) -- node[right] {$0$} (2);
        \draw[->, >=latex, line width=0.5pt] (S) -- (1);
        \draw[->, >=latex, line width=0.5pt] (S) -- (2);
        \draw[->, >=latex, line width=0.5pt] (1) -- (T);
        \draw[->, >=latex, line width=0.5pt] (2) -- (T);
    \end{tikzpicture}
\end{center}
This game is a variant of the well-known Braess's paradox \cite{braess1968paradoxon}. In this game, both players want to go from $S$ to $T$. The costs of the edges are annotated on the graph where $c$ denotes the number of players going through the edge. Let $\Sigma$ denote the strategy profile where both players go through $S\to 1\to 2\to T$. We can see that $\Sigma$ is a Nash Equilibrium if and only if $x$ and $y$ are in different coalitions. This is exactly what we have in \cref{lem:normal_form_gadget}. Moreover, the products of congestion games can also be represented as a congestion game. Therefore, we can use this gadget to replace $\NF(x,y)$ in \cref{alg:normal_form_games} and solve the CSL problem with congestion games. The sample complexity upper bound of this algorithm is $n \log_2 n + 3n$ as well.

\paragraph{CSL with graphical games.} A graphical game \cite{kearns2001graphical} is represented by a graph $G$, where each vertex denotes a player. There is an edge between a pair of vertices $x$ and $y$ if and only if their utilities are dependent on each other's strategy. To limit the size of the representation of a graphical game, a common way is to limit the maximum vertex degree $d$ in $G$. We show that with a slight modification, IG can be extended to solve the CSL problem with graphical games of maximum vertex degree $d = 1$ with the same sample complexity upper bound $n \log_2 n + 3n$. The details are deferred to \cref{app:csl_with_graphical_games}.

\section{CSL with Auctions}
\label{sec:csl_with_auctions}

We now pivot from classic games to a more practical class of games: second-price auctions with personalized reserves \cite{paes2016field}. Collusion of multiple agents in auctions has already been extensively observed (see, e.g., \citeauthor{milgrom2004putting} \citeyear{milgrom2004putting}). The auction mechanisms can be exploited if these coordinated bidders deviate simultaneously. Thus, it is important to study the CSL problem with auctions. We refer to this variant of CSL as AuctionCSL.

In such an auction, each agent $i$ has a private value $v_i$ for the item being auctioned and a personalized reserve price $r_i$. Each agent $i$ submits a bid $b_i$ to the auction, after which the auction will choose an agent with the highest bid $i^*\in\arg\max_{i\in \N}\{b_i\}$ and offer the item to $i^*$ with price $p=\max\{r_{i^*},\max_{i\ne i^*}\{b_i\}\}$. The agent $i^*$ can choose to accept or reject the offer. If $i^*$ rejects, the auction ends with no transaction. Otherwise, $i^*$ pays $p$ and gets the item. The item is then redistributed within $i^*$'s coalition to the agent with the highest private valuation for maximum coalition utility.

In this section, we consider an online auction setting where we play the role of an auctioneer. Our goal is to recover the coalition structure $\S^*$.
As the values $\{v_i\}$ are determined by each agent's valuation for the item being auctioned, we study the setting where we can only design the reserve prices $\{r_i\}$. We assume that a stream of items will arrive to be auctioned, whose values $\{v_i\}$ are randomly generated each time, and we have no power to design them. However, we assume that we know $\{v_i\}$ before designing the reserve prices $\{r_i\}$: this happens when we are sufficiently acquainted with the agents, so we can estimate their values given a certain item. 
We fix the default strategy profile of the agents as truthful bidding, i.e., $b_i=v_i$ for all $i\in\N$.

\subsection{Group Testing via Auction Gadgets}
\label{subsec:gadget_construction_auction}

Inspired by \cref{alg:normal_form_games}, we still would like a way to tell whether there is an agent $j$ inside a set $T$ that is in the same coalition with another agent $i$ using the result of a single auction. 
However, since the product of two auctions is no longer an auction (see \cref{appsub:auctions_are_not_closed_under_product}), the same method as in \cref{sec:csl_with_normal_form_games} is not appropriate. Therefore, we need to design a new gadget for auctions. The main idea remains the same.

\begin{definition}
    \label{def:auction_gadget}
    Let $\v \in [0,1]^n$ and $T \subseteq \N$. An \textbf{auction gadget} $\A(\v,T)$ is a second price auction with personalized reserves where the values of the agents are $\v$. Let $v_{\textup{max}}$ and $v_{\textup{smax}}$ be the maximum and second maximum value among $\v$ respectively. The reserve prices of the agents are defined as
    \begin{equation*}
        r_i=\left\{\begin{array}{ll}
            v_{\textup{smax}} & (i \in T) \\
            v_{\textup{max}} & (i \notin T).
        \end{array}\right.
    \end{equation*}
    The \textbf{default strategy profile} is bidding $b_i=v_i$ for all $i\in\N$.
\end{definition}

We similarly establish the following connection between the result of an auction gadget and the coalition structure.

\begin{lemma}
    \label{lem:auction_gadget}
    Let $\v \in [0, 1]^n$ be a vector such that $v_{i}$ is the unique maximum.  
    Let $T\subseteq \N \backslash \{ i\}$. Then bidding truthfully in $\A(\v,T)$ is a Nash Equilibrium if and only if $[i]_{\S^*} \neq [j]_{\S^*}$ (i.e., $i$ and $j$ are in different coalitions) for all $j\in T$.
\end{lemma}

\begin{proofof}{\cref{lem:auction_gadget}}
    Let $v_{\textup{max}}=v_i$, $v_{\textup{smax}}=\max_{j\ne i}\{v_j\}$.
    If $\exists j\in T$, such that $i$ and $j$ are in the same coalition, then they can jointly deviate by bidding $b_{i}=v_{\textup{smax}}$ and $b_{j}=v_{\textup{max}}$. In this way, $j$ wins the auction with price $p=v_{\textup{smax}}$. $j$ can accept the item with this price, and redistribute it to $i$. The total utility of $i$ and $j$'s coalition increases from $0$ to $v_{\textup{max}}-v_{\textup{smax}}$. Thus bidding truthfully is not a Nash Equilibrium. If $\forall j\in T$, $i$ and $j$ are in different coalitions, then (i) the unilateral deviation of $i$'s coalition cannot lead to positive utility as all members in this coalition have a reserve price of $v_{\textup{max}}$ (ii) the unilateral deviation of any other coalitions cannot lead to positive utility as the maximum value among them is $v_{\textup{smax}}$, and the reserve prices of them is at least $v_{\textup{smax}}$. Thus bidding truthfully is a Nash Equilibrium.
\end{proofof}



From \cref{lem:auction_gadget} and \cref{lem:product_of_normal_form_game_gadget}, we can see that auction gadgets are analogous to normal form gadgets. Assuming we have the freedom to design valuation vectors $\v\in[0, 1]^n$ for an auctioned item, then $\A(\v, T)$ may be used to determine if an agent in $T$ that is also in $[i]_{\S^*}$. This yields an algorithm similar to IG (\cref{alg:normal_form_games}) for solving AuctionCSL under this simplifying assumption. We describe this algorithm and its theoretical guarantees in \cref{app:csl_with_designed_auctions}.

\subsection{IG under Auctions with Random Valuations}
\label{subsec:ig_with_random_values}

In real auctions, valuations of items are beyond our control. We model this more realistic setting by assuming that the values are drawn from an item pool $\V$, which is a distribution $\U[0,1]^n$ over $\mathbb R^n$. 
Intuitively, the randomness of the values makes CSL in this setting significantly more challenging than the normal form game setting, as we cannot guarantee progress of the algorithm if we get an unlucky draw of the values. For example, if the item has $0$ value for all agents, then truthful bidding will always be a Nash Equilibrium no matter what the reserve prices are. This suggests that we can at best hope for a guarantee on the expected sample complexity.
We design AuctionIG for this setting.


\begin{algorithm}[htbp]
    \caption{IG with Auctions (AuctionIG)}
    \label{alg:auction_random_values}

    \textbf{Input}: The number of agents $n$ and an observation oracle $\O$

    \textbf{Output}: A coalition structure $\S$ of the agents

    \begin{algorithmic}[1] 
        \STATE Let $\S\gets\{\{1\},\{2\},\dots,\{n\}\}$.
        \STATE Let $T_{i} \gets \varnothing$ for all $i\in\N$.
        \STATE Let $C_{i} \gets 0$ for all $i\in\N$.
        \STATE Let $T_{\textup{finalized}} \gets \varnothing$.
        \WHILE {$T_{\textup{finalized}} \ne \N$}
            \STATE Get $\v \sim \V$.
            \STATE Let $x \gets \arg\max_{i\in\N}\{v_i\}$ and $C_x \gets C_x + 1$.
            \IF {$T_x = \varnothing$}
                \IF {$\O(\A(\v,\N\setminus[x]_\S)) = \textup{false}$}
                    \STATE Let $T_i \gets \N\setminus[x]_\S$ for all $i \in [x]_\S$.
                \ELSE
                    \STATE Let $T_{\textup{finalized}} \gets T_{\textup{finalized}} \cup [x]_\S$.
                \ENDIF
            \ELSE
                \STATE Partition $T_x$ into $T_\alpha,T_\beta$ where $\left||T_\alpha|-|T_\beta|\right|\leq 1$.
                \IF {$\O(\A(\v,T_\alpha)) = \textup{false}$}
                    \STATE Let $T_i \gets T_\alpha$ for all $i \in [x]_\S$.
                \ELSE
                    \STATE Let $T_i \gets T_\beta$ for all $i \in [x]_\S$.
                \ENDIF
            \ENDIF
            \IF {$|T_x| = 1$}
                \STATE Let $y\gets \textup{the only element in $T_x$}$.
                \STATE Merge $[x]_\S$ and $[y]_\S$ in $\S$.
                \STATE Let $T_{i} \gets \varnothing$ for all $i \in [x]_\S$.
            \ENDIF
        \ENDWHILE
        \STATE \textbf{return} $\S$.
    \end{algorithmic}
\end{algorithm}

The main idea of AuctionIG is still similar to IG (\cref{alg:normal_form_games}). For agent $x$, we try to iteratively find other agents in $x$'s coalition $[x]_{\S^*}$ using binary search. However, as we do not have control over which agent has the largest value, we cannot do this sequentially for each agent as in IG. Instead, we run multiple instances of binary search in parallel, each progressing depending on which item is drawn.

In AuctionIG, for each $i \in N$ we maintain $T_i$ as a set containing another agent in $i$'s coalition (Line 2), $C_i$ as the number of times $v_i$ has appeared as the largest value in $\v$ (Line 3), and $T_{\textup{finalized}}$ as the set of agents whose coalitions have been finalized (Line 4). Each time we draw an item $\v$ from $\V$, we find the agent $x$ with the largest value (Lines 6 to 7), and try to proceed with the binary search to expand $x$'s coalition. If $T_x = \varnothing$, then we should start a new binary search for $x$'s coalition (Lines 9 to 12). We first check whether there is an agent in $x$'s coalition in $\N\setminus[x]_\S$. If so, we set $T_i$ to $\N\setminus[x]_\S$ for all $i \in [x]_\S$; otherwise, we know that $x$'s coalition is finalized, and we add the entire coalition to $T_{\textup{finalized}}$ (Line 12). 
If $T_x\ne\varnothing$, then we are in the middle of a binary search for $x$'s coalition (Lines 13 to 18). We partition $T_x$ into $T_\alpha$ and $T_\beta$ and check whether there is an agent in $x$'s coalition in $T_\alpha$. If so, we set $T_i$ to $T_\alpha$ for all $i \in [x]_\S$; otherwise, we set $T_i$ to $T_\beta$ for all $i \in [x]_\S$. If $|T_x|=1$, then we have found another agent $y$ in $x$'s coalition (Lines 20 to 22). We merge their coalitions and set $T_i$ to $\varnothing$ for all $i \in [x]_\S$, indicating that binary search should be restarted for this coalition. The outermost loop runs until $T_{\textup{finalized}} = \N$, which means that we have finalized the coalitions of all agents.

To analyze AuctionIG, we utilize the invariants in the following Lemma, whose proof is deferred to \cref{appsub:proof_of_lem_state}.

\begin{restatable}{lemma}{state}
    \label{lem:state}
    Let $\S^*$ be the correct coalition structure. The following holds throughout the execution of AuctionIG. 
    \begin{enumerate}[(a)]
        \item $[i]_\S \subseteq [i]_{\S^*},\forall i\in \N$.
        \item $[i]_\S = [i]_{\S^*},\forall i\in T_{\textup{finalized}}$.
        \item $T_i=T_j \textup{ if } [i]_\S=[j]_\S$.
        \item $\exists j\in T_i \textup{ such that } j\in[i]_{\S^*}\setminus[i]_\S \textup{ if } T_i \ne \varnothing$.
    \end{enumerate}
\end{restatable}

Next, we show a termination condition for AuctionIG.

\begin{restatable}{lemma}{termination}
    \label{lem:termination}
    AuctionIG terminates no later than the time when $C_i\geq 2\log_2n + 4$ holds for all $i\in\N$.
\end{restatable}

We will prove \cref{lem:termination} in \cref{appsub:proof_of_lem_termination}. To sketch the proof, we define $\S_i = \{[j]_\S\mid j\in [i]_{\S^*}\}$ and
\begin{equation*}
    f(T)=\left\{\begin{array}{ll}
        \lceil\log_2 n\rceil + 1 & (T = \varnothing) \\
        \lceil\log_2 |T|\rceil & (T \ne \varnothing).
    \end{array}\right.
\end{equation*}
According to \cref{lem:state} (c), we can unambiguously use $T_S$ to denote $T_x$ for any $x\in S\in \S_i$ and define the potential function $\Phi_i(\mathbf T, \S)=\lceil \log_2 n\rceil \cdot |\S_i| + \sum_{S\in\S_i} f(T_S)$, where $\mathbf T$ is the vector of all $T_i$. Intuitively, the potential function characterizes the remaining progress associated with agent $i$, where $\lceil \log_2 n\rceil \cdot |\S_i|$ adds $\lceil \log_2 n\rceil$ for each unmerged coalition in $\S_i$ and $\sum_{S\in\S_i} f(T_S)$ adds $\lceil \log_2 |T_S|\rceil$ for each coalition $S\in\S_i$, indicating the remaining steps in the binary search. To complete the proof, we show that $\Phi_i(\mathbf T, \S)$ decreases by at least $1$ after any $C_j$ for $j\in[i]_{\S^*}$ increases. 

\cref{lem:termination} shows that we will have finalized the coalition structures when we have gotten for each agent $i$, $2\log_2n + 4$ items that are most valuable to $i$.
This connects the sample complexity of AuctionIG to a well-studied problem in statistics, the coupon collector's problem \cite{newman1960double,erdHos1961classical}. In this problem, there are $n$ types of coupons, and each time we draw a coupon, we get a coupon of a uniformly random type. We want to collect $k$ sets of coupons, where each set contains one coupon of each type. The coupon collector's problem asks for the expected number of draws needed to collect $k$ sets of coupons $T_{\textup{ccp}}(n,k)$. 

\cref{lem:termination} demonstrates that the sample complexity of AuctionIG is upper bounded by $T_{\textup{ccp}}(n, 2\log_2 n + 4)$. Combining this with the result of \citet{papanicolaou2020old} from the coupon collector's problem's literature, we have \cref{thm:auction_random_values} with its proof deferred to \cref{appsub:proof_of_thm_auction_random_values}.

\begin{restatable}{theorem}{auctionrandomvalues}
    \label{thm:auction_random_values}
    AuctionIG solves AuctionCSL with \textbf{expected} sample complexity upper bounded by $(4.16 + o(1))n \log_2 n$.
\end{restatable}

Using Markov's inequality, we can also transform \cref{thm:auction_random_values} into a PAC learning type of result as below.

\begin{corollary}[PAC Complexity]
    \label{cor:auction_random_values}
    For any $\delta\in(0,1)$, AuctionIG correctly learns the coalition structure with probability at least $1-\delta$ using $(4.16 + o(1))\frac{n \log_2 n}{\delta}$ auctions.
\end{corollary}

We also study the performance of AuctionIG in the special cases when $m=1$ and $m=n$, i.e., when there is only one coalition and when each agent is in a separate coalition. The proof is given in \cref{appsub:proof_of_thm_auction_random_special_case}.

\begin{restatable}{theorem}{auctionrandomspecialcase}
    \label{thm:auction_random_special_case} Let $m$ be the number of coalitions.
    \begin{enumerate}[(a)]
        \item When $m=1$, the sample complexity of AuctionIG is bounded by $2n\log_2n+4n$ determinsitically.
        \item When $m=n$, the \textbf{expected} sample complexity of AuctionIG is exactly $nH_n\leq (0.70 + o(1))n \log_2 n$.
    \end{enumerate}
\end{restatable}

\section{Experiments}
\label{sec:experiments}

\begin{figure*}[t]
	\centering

	\begin{subfigure}{0.32\textwidth}
	  \centering
	  \includegraphics[width=\linewidth]{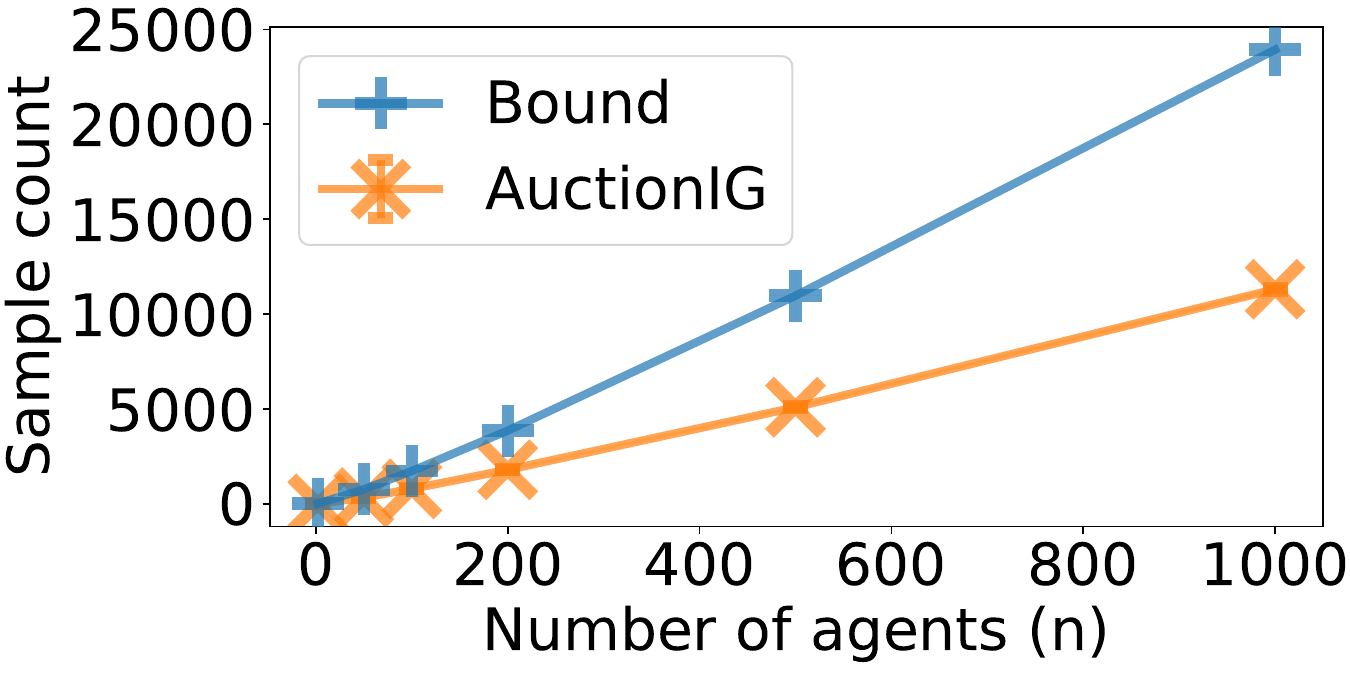}
	  \caption{Expected sample complexity $(m = 1)$}
	  \label{subfigure:expected_m=1}
	\end{subfigure}
	\hfill
	\begin{subfigure}{0.32\textwidth}
	  \centering
	  \includegraphics[width=\linewidth]{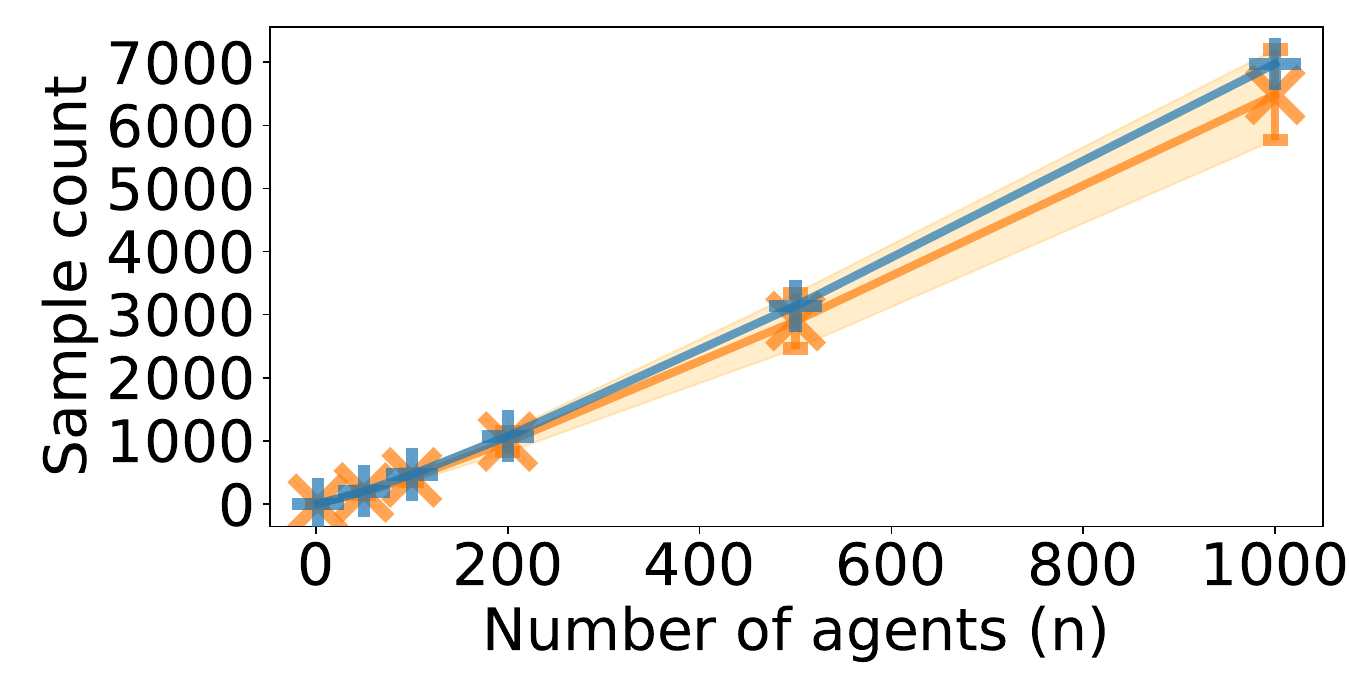}
	  \caption{Expected sample complexity $(m = n)$}
	  \label{subfigure:expected_m=n}
	\end{subfigure}
	\hfill
	\begin{subfigure}{0.32\textwidth}
	  \centering
	  \includegraphics[width=\linewidth]{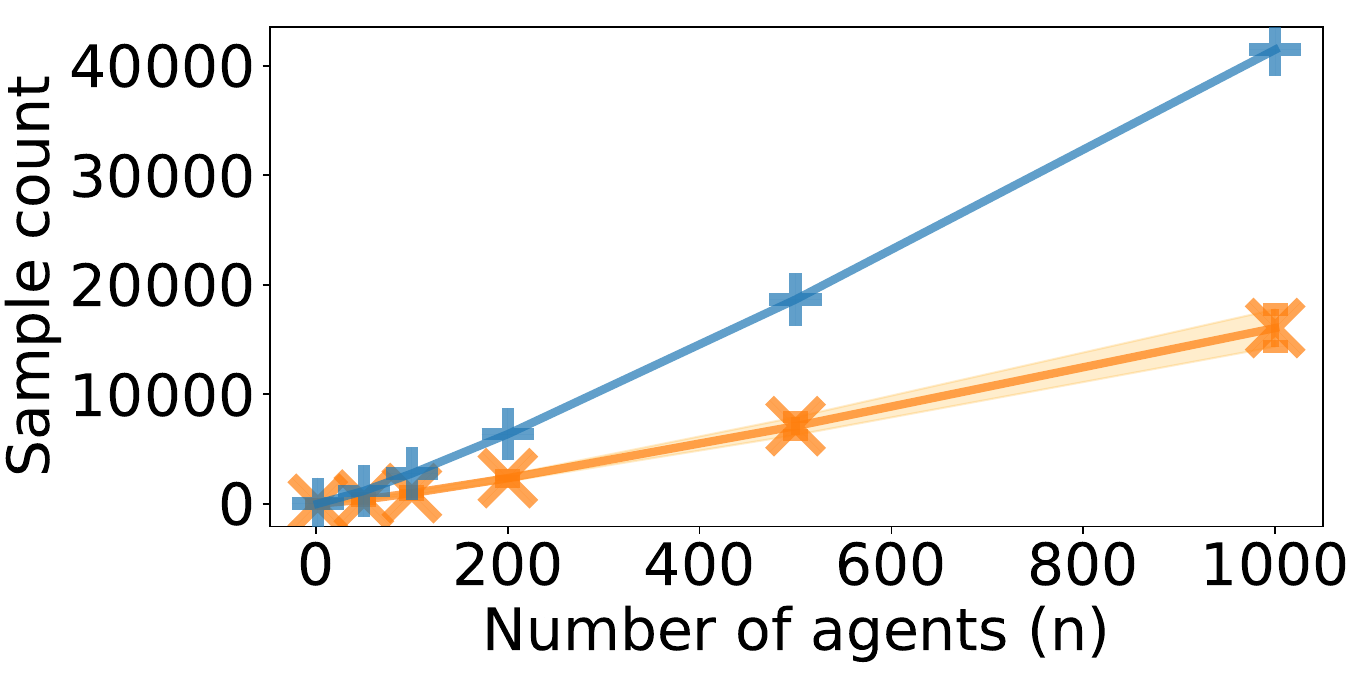}
	  \caption{Expected sample complexity $(m \sim \U[n])$}
	  \label{subfigure:expected_random_m}
	\end{subfigure}

    \begin{subfigure}{0.32\textwidth}
        \centering
        \includegraphics[width=\linewidth]{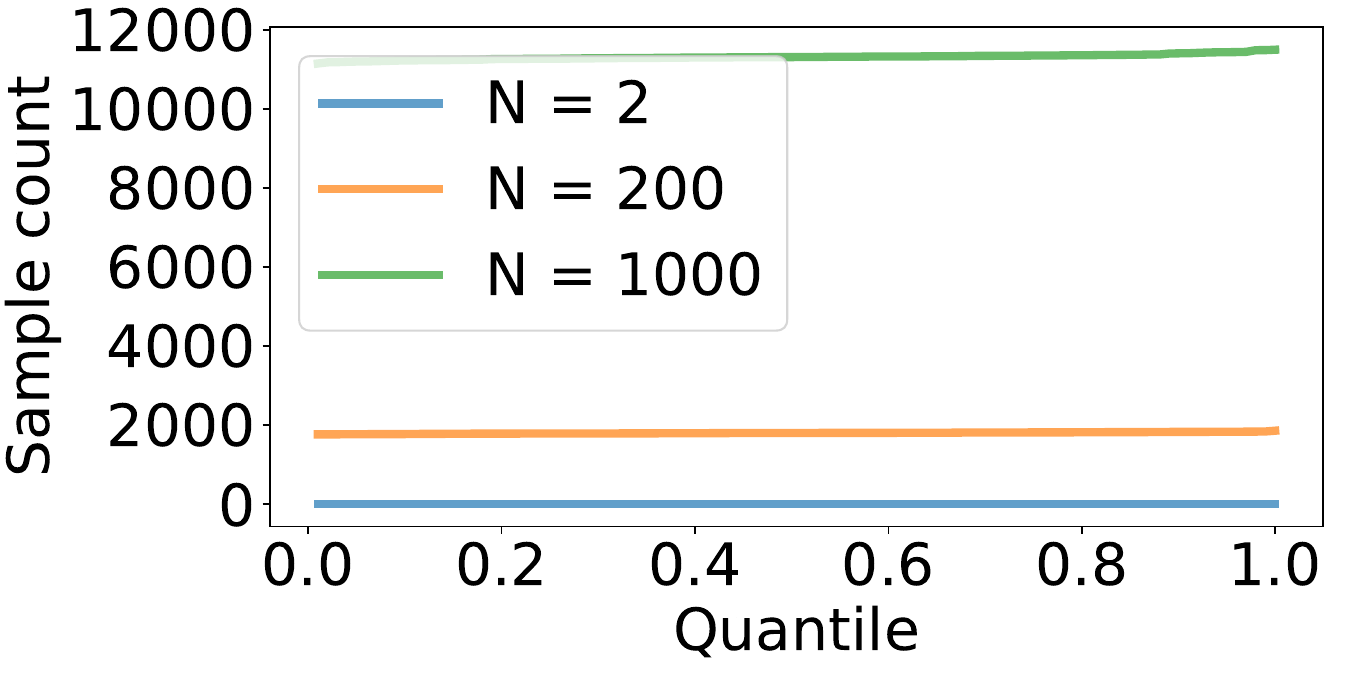}
        \caption{Sample complexity CDF $(m = 1)$}
        \label{subfigure:cdf_m=1}
      \end{subfigure}
      \hfill
      \begin{subfigure}{0.32\textwidth}
        \centering
        \includegraphics[width=\linewidth]{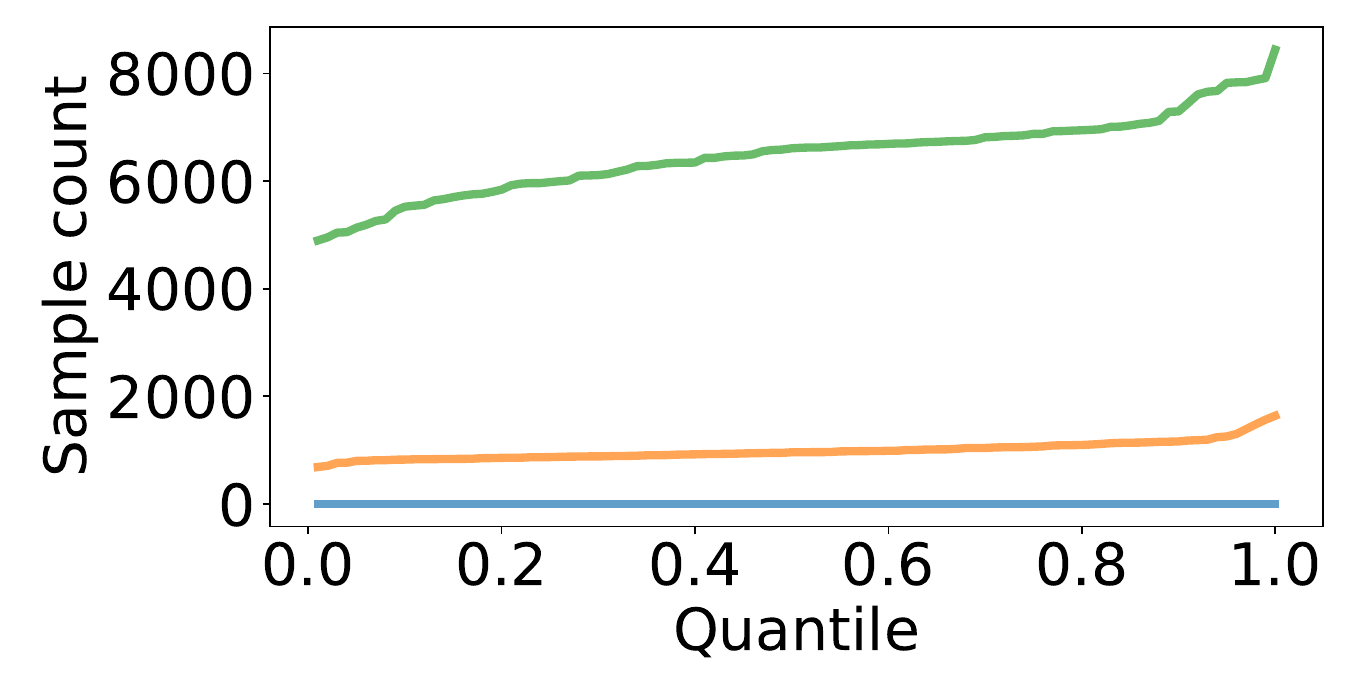}
        \caption{Sample complexity CDF $(m = n)$}
        \label{subfigure:cdf_m=n}
      \end{subfigure}
      \hfill
      \begin{subfigure}{0.32\textwidth}
        \centering
        \includegraphics[width=\linewidth]{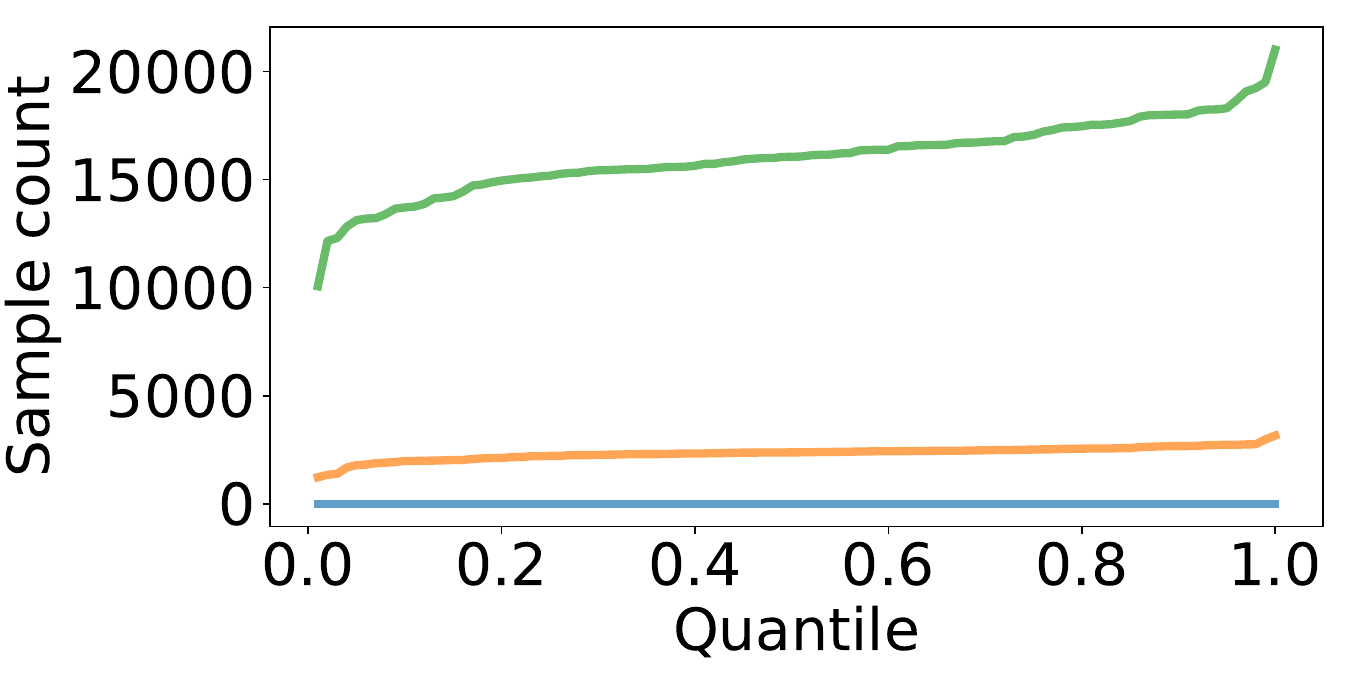}
        \caption{Sample complexity CDF $(m \sim \U[n])$}
        \label{subfigure:cdf_random_m}
      \end{subfigure}

      \begin{subfigure}{0.32\textwidth}
        \centering
        \includegraphics[width=\linewidth]{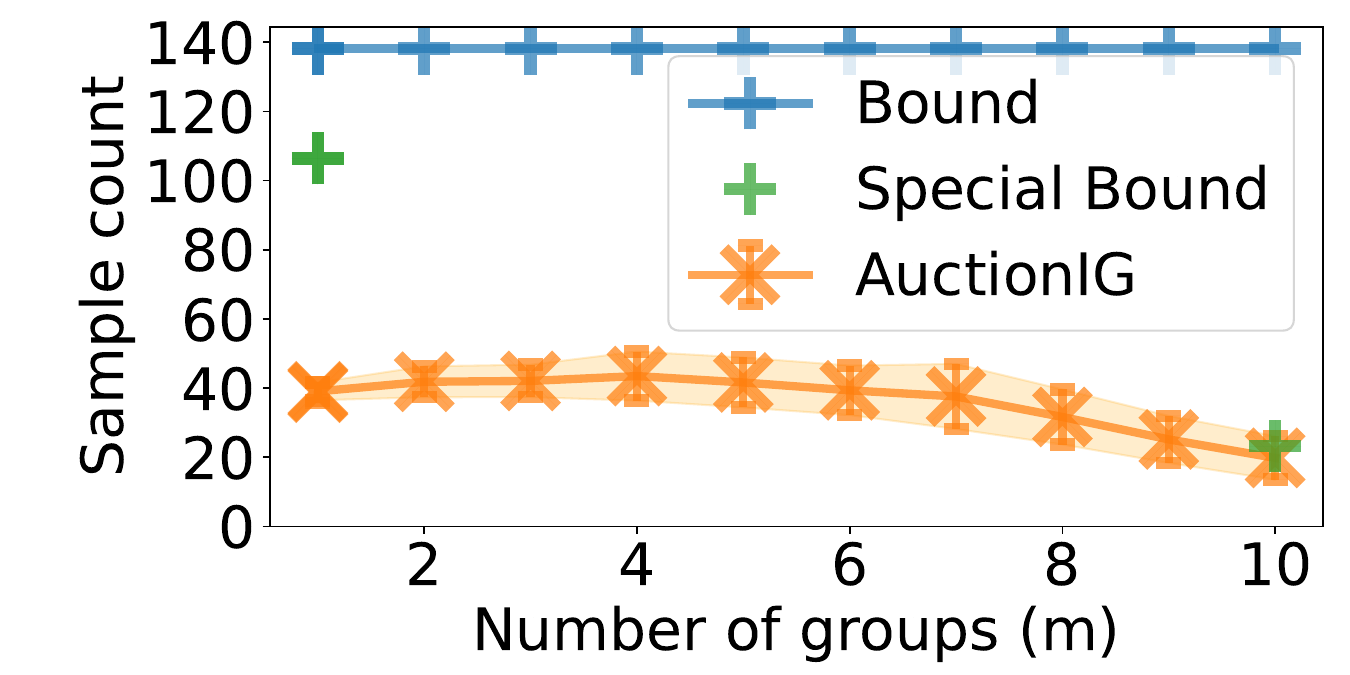}
        \caption{Expected sample complexity $(n = 10)$}
        \label{subfigure:expected_n=10}
      \end{subfigure}
      \hfill
      \begin{subfigure}{0.32\textwidth}
        \centering
        \includegraphics[width=\linewidth]{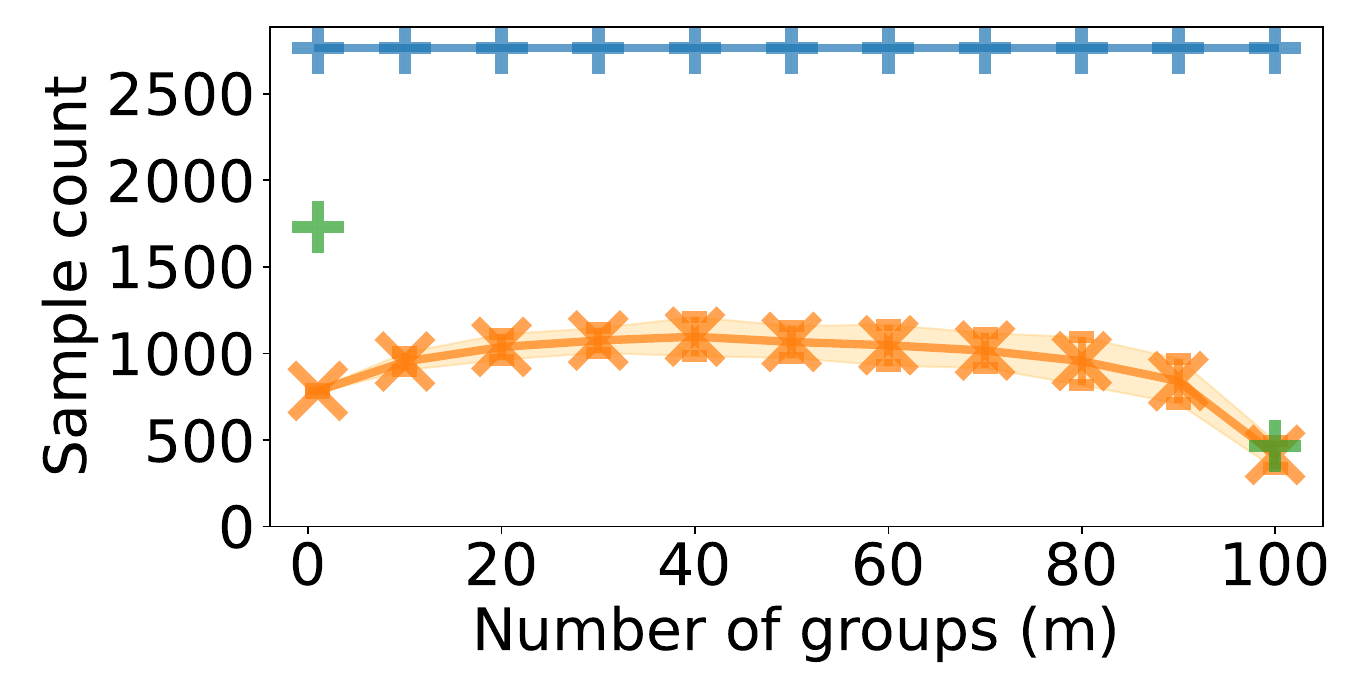}
        \caption{Expected sample complexity $(n = 100)$}
        \label{subfigure:expected_n=100}
      \end{subfigure}
      \hfill
      \begin{subfigure}{0.32\textwidth}
        \centering
        \includegraphics[width=\linewidth]{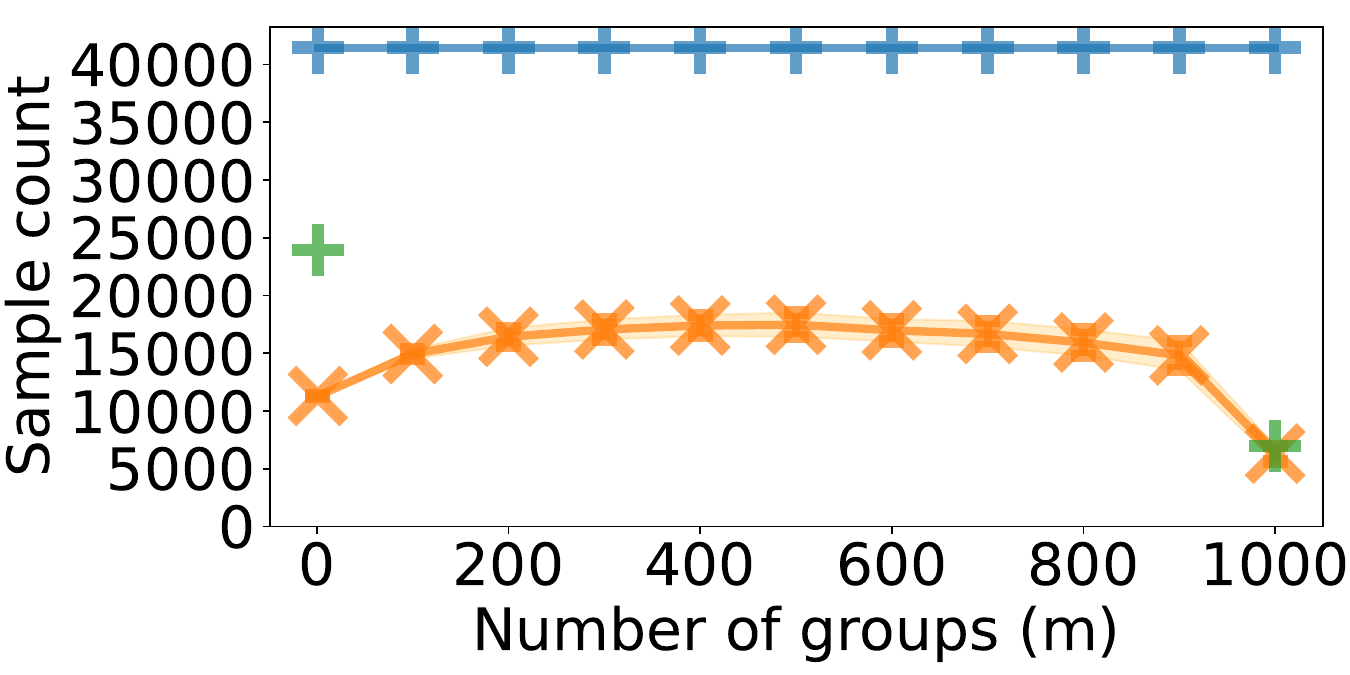}
        \caption{Expected sample complexity $(n = 1000)$}
        \label{subfigure:expected_n=1000}
      \end{subfigure}

	\caption{Performance of AuctionIG under different settings of $n$ and $m$. \textit{Bound} refers to the theoretical bound of its expected sample complexity given by \cref{thm:auction_random_values} and whenever applicable \cref{thm:auction_random_special_case}. \textit{AuctionIG} refers to the average sample complexity of AuctionIG over 100 runs with error bars indicating its standard deviation. The results show that the actual performance of AuctionIG is always within a constant factor of its theoretical bounds given in \cref{thm:auction_random_values,thm:auction_random_special_case}.}
	\label{fig:main}
\end{figure*}

We conduct experiments to evaluate the performance of our algorithms in practice. As IG (\cref{alg:normal_form_games}, normal form games) is deterministic and theoretically optimal (up to low order terms) in sample complexity, we only evaluate AuctionIG (\cref{alg:auction_random_values}, auctions). We implement it in Python and evaluate it on a server with 56 cores and 504G RAM, running Ubuntu 20.04.6. The source codes can be found at \url{https://github.com/YixuanEvenXu/coalition-learning}.

\paragraph{Experiment setup.} We evaluate AuctionIG under different settings of $n$ and $m$, where $n$ is the number of agents and $m$ is the number of coalitions. For each setting, we fix $n$ and either fix $m$ or sample $m$ from $\U[n]$. Then, we synthesize a coalition structure $\S^*$ with exactly $n$ agents and $m$ coalitions at random. We then run AuctionIG, check the correctness of its output, and record the sample complexity (the total number of samples used). We repeat this process 100 times and report the distribution of the sample complexity. We also report the theoretical upper bound of the expected sample complexity given by \cref{thm:auction_random_values} and whenever applicable \cref{thm:auction_random_special_case}. The results are shown in \cref{fig:main}.

\paragraph{AuctionIG's performance with different $n$.} As shown in \cref{subfigure:expected_m=1,subfigure:expected_m=n,subfigure:expected_random_m}, we let $n = \{2, 50, 100, 200, 500, 1000\}$ and consider fixing $m = 1$ (\cref{subfigure:expected_m=1}), fixing $m = n$ (\cref{subfigure:expected_m=n}) and sampling $m$ from $\U[n]$ (\cref{subfigure:expected_random_m}). For $m = 1,n$, we apply the bounds given in \cref{thm:auction_random_special_case}, and for $m \sim \U[n]$, we apply the bound given in \cref{thm:auction_random_values}. The results show that the actual performance of AuctionIG is always within a constant factor of its theoretical bounds given in \cref{thm:auction_random_values,thm:auction_random_special_case}. Moreover, when $m = n$, the actual performance is very close to the theoretical bound.

\paragraph{AuctionIG's performance with different $m$.} As shown in \cref{subfigure:expected_n=10,subfigure:expected_n=100,subfigure:expected_n=1000}, we let $m = \{1, 0.1n, 0.2n, \dots, n\}$ and consider fixing $n = 10$ (\cref{subfigure:expected_n=10}), $n = 100$ (\cref{subfigure:expected_n=100}) and $n = 1000$ (\cref{subfigure:expected_n=1000}). We plot the theoretical bounds given in \cref{thm:auction_random_values} for all $m$ and those given in \cref{thm:auction_random_special_case} for $m = 1,n$. The results show that when $m \in (1, n)$, the sample complexities of AuctionIG are similar across different values of $m$. However, when $m = 1, n$, the sample complexities are significantly lower. This trend is increasingly visible when $n$ grows larger. This shows that \cref{thm:auction_random_special_case} complements \cref{thm:auction_random_values} well in the sense that it provides a tighter bound for the special cases when $m = 1, n$.

\begin{table}[htbp]
	\centering
	\begin{tabular}{|c|c|c|c|c|}
		\hline
		\multicolumn{2}{|c|}{Correct Probability} & $50\%$ & $90\%$ & $99\%$ \\
		\hline
        \multicolumn{2}{|c|}{Bound in \cref{cor:auction_random_values}} & 82916 & 414577 & 4145767 \\
        \hline
        \multirow{3}{*}{Algorithm} & $m = 1$ & 11306 & 11401 & 11482 \\
        \cline{2-5}
        & $m = n$ & 6610 & 7294 & 7915 \\
        \cline{2-5}
        & $m \sim \U[n]$ & 16055 & 18009 & 19510 \\
		\hline
	\end{tabular}
	\caption{The empirical sample complexity of AuctionIG with $50\%$, $90\%$ and $99\%$ probability of correctness when $n = 1000$ and the corresponding bounds given in \cref{cor:auction_random_values}. The results show that in practice AuctionIG performs much better than the bounds given in \cref{cor:auction_random_values}.}
	\label{table:pac}
\end{table}

\paragraph{PAC complexity of AuctionIG.} As shown in \cref{subfigure:cdf_m=1,subfigure:cdf_m=n,subfigure:cdf_random_m}, we evaluate the PAC complexity of AuctionIG by plotting the CDFs of its sample complexity over 100 runs under different settings of $n$ and $m$. We also highlight several points on the CDFs that correspond to the sample complexity of AuctionIG with $50\%$, $90\%$, and $99\%$ probability of correctness when $n = 1000$ in \cref{table:pac}. We can see that the actual sample complexity of AuctionIG is relatively stable across different runs and is much lower than the theoretical bounds given in \cref{cor:auction_random_values} when we require a high probability of correctness. This is because \cref{cor:auction_random_values} is derived using Markov's inequality, which is a very loose bound. In fact, with a finer-grained analysis of the coupon collector's problem, we can improve it using the limit distribution of the coupon collector's problem (see e.g. \citeauthor{papanicolaou2020old} \citeyear{papanicolaou2020old}). However, in that way, we will not be able to write the PAC complexity in a simple closed form.

\paragraph{Summary of experiment results.} The experiments show that \cref{thm:auction_random_values,thm:auction_random_special_case} characterize the expected sample complexity of AuctionIG well with a tight constant, especially when $m = n$ where the bounds are almost perfect. Moreover, the empirical PAC complexity of AuctionIG is much lower than the bounds given in \cref{cor:auction_random_values}, demonstrating its practicality.

\section{Conclusion and Discussion}
\label{sec:conclusion_and_discussion}
In this paper, we propose and study the Coalition Structure Learning (CSL) and AuctionCSL problems under the one-bit observations. We present a novel Iterative Grouping (IG) algorithm and its counterpart AuctionIG to efficiently tackle these problems, both achieving a sample complexity with asymptotically matching lower bounds. Empirical results demonstrate that these algorithms are indeed sample efficient and useful in practice. Future work includes (i) handling cases where players are aware of and are strategically manipulating our algorithm, (ii) handling bounded rationality, (iii) more general classes of observations, and (iv) admitting equilibrium concepts beyond Nash.

\newpage

\section*{Acknowledgements}

This work was supported in part by NSF grant IIS-2046640 (CAREER), IIS-2200410, and Sloan Research Fellowship.
Additionally, we thank Davin Choo and Hanrui Zhang for helpful discussions about this work.

\bibliography{ref}

\begin{table*}[htbp]
	\centering
	\begin{tabular}{|c|c|l|}
		\hline
		Introduced & Notation & Meaning \\
		\hline
        \multirow{10}{*}{\cref{sec:introduction}} & $n$ & The number of agents \\
        \cline{2-3}
        & $m$ & The number of coalitions \\
        \cline{2-3}
        & $\N$ & The set of all agents $\{1,2,\dots,n\}$ \\
        \cline{2-3}
        & $i,j,x,y$ & Agents within the set $\N$ \\
        \cline{2-3}
        & $S$ & A coalition \\
        \cline{2-3}
        & $\S$ & A coalition structure \\
        \cline{2-3}
        & $[i]_\S$ & The coalition containing $i$ in $\S$ \\
        \cline{2-3}
        & $\G$ & A game \\
        \cline{2-3}
        & $\O$ & The observation oracle \\
        \cline{2-3}
        & $\Sigma$ & A mixed strategy profile \\
        \hline
        \multirow{7}{*}{\cref{sec:csl_with_normal_form_games}} & $\S^*$ & The ground truth coalition structure \\
        \cline{2-3} 
        & $T, T_\alpha, T_\beta$ & Sets of agents \\
        \cline{2-3}
        & $(\G,\Sigma)$ & A game-strategy pair (\cref{def:game-strategy-pair}) \\
        \cline{2-3}
        & $\NF(x,y)$ & A normal form gadget (\cref{def:normal_form_gadget}) \\
        \cline{2-3}
        & $\sigma_1\times\sigma_2$ & The product of mixed strategies (\cref{def:product_of_mixed_strategies}) \\
        \cline{2-3}
        & $(\G_1,\Sigma_1)\times (\G_2,\Sigma_2)$ & The product of game-strategy pairs (\cref{def:product_of_normal_form_games}) \\
        \cline{2-3}
        & $\prod_{\theta}(\G_\theta,\Sigma_\theta)$ & The product of multiple game-strategy pairs \\
        \hline
        \multirow{7}{*}{\cref{sec:csl_with_auctions}} & $b_i$ & The bid of agent $i$ \\
        \cline{2-3} 
        & $\mathbf b,\{b_i\}$ & The bids of all agents \\
        \cline{2-3}
        & $v_i$ & The value of agent $i$ \\
        \cline{2-3}
        & $\v,\{v_i\}$ & The valuse of all agents \\
        \cline{2-3}
        & $r_i$ & The reserve price of agent $i$ \\
        \cline{2-3}
        & $\r,\{r_i\}$ & The reserve prices of all agents \\
        \cline{2-3}
        & $\A(\v,T)$ & An auction gadget (\cref{def:auction_gadget}) \\
        \hline
	\end{tabular}
	\caption{List of key notations.}
	\label{table:notations_csl}
\end{table*}
\newpage
\appendix

\section{List of Notations}
\label{app:list_of_notations}

We summarize key notations in this paper in \cref{table:notations_csl}.

\section{Missing Proofs in \cref{sec:csl_with_normal_form_games,sec:csl_with_auctions}}
\label{app:missing_proofs}

\subsection{Proof of \cref{thm:normal_form_games}}
\label{appsub:proof_of_thm_normal_form_games}

\normalformgames*

\begin{proofof}{\cref{thm:normal_form_games}}
    We first prove the correctness of \cref{alg:normal_form_games}. Let the ground truth coalition structure of the agents be $\S^*$. We will show that after the $i$-th iteration of the outer for loop, $[j]_\S=[j]_{\S^*},\forall j\leq i$ and $[j]_\S\subseteq[j]_{\S^*},\forall j > i$. Below, we prove this statement by induction.
    
    The base case $(i=0)$ is trivial by the initialization of $\S$ on Line 1. Suppose the statement holds for $i-1$. By \cref{lem:product_of_normal_form_game_gadget}, if $\O(\prod_{[j]_\S \ne [i]_\S}\NF(i,j)) = \textup{true}$ on Line 3, then no agent outside $[i]_\S$ is in the same coalition with $i$. This means that $[i]_\S=[i]_{\S^*}$ already. Thus the statement holds for $i$ in this case. If $\O(\prod_{[j]_\S \ne [i]_\S}\NF(i,j)) = \textup{false}$ on Line 3, then there must be some agent in $T=\{j\in\N\mid [j]_\S \ne [i]_\S\}$ that is in the same coalition with $i$. The while loop in Lines 5 to 10 then splits $T$ into $2$ non-empty subsets $T_\alpha$ and $T_\beta$. If $T_\alpha$ contains an agent in the same coalition with $i$, then $\O(\prod_{j\in T_\alpha}\NF(i,j)) = \textup{false}$, and $T$ is set to $T_\alpha$. Otherwise, $T_\beta$ must contain an agent in the same coalition with $i$, and $T$ is then set to $T_\beta$. In either case, $T$ becomes smaller while still containing an agent in the same coalition with $i$. This process will terminate when $T$ contains only one agent $j$, and $j$ must be in the same coalition with $i$. We merge the coalitions of $i$ and $j$ on Line 12. The while loop in Lines 3 to 12 repeats this process until $[i]_\S=[i]_{\S^*}$. Also, as we only merge agents within the same coalition in $\S^*$, $[j]_\S\subseteq[j]_{\S^*},\forall j > i$ still holds after the merge. Thus the statement holds for $i$. Therefore, inductively, the statement holds for all $i = \{0,1,\dots,n\}$. The correctness of \cref{alg:normal_form_games} then follows from the statement when $i = n$.

    Next, we prove the sample complexity of \cref{alg:normal_form_games}. The outer for loop runs for $n$ iterations. In the $i$-th iteration, suppose the body of the while loop in Lines 3 to 12 runs for $k_i$ times. Then, as the inner while loop in Lines 5 to 10 terminates in $\left\lceil \log_2 n\right\rceil$ iterations, the total number of oracle accesses should be upper bounded by $\sum_{i=1}^{n}(k_i\left\lceil \log_2 n\right\rceil + k_i + 1)$. Note that each time the while loop in Lines 3 to 12 runs, the number of coalitions in $\S$ decreases by one as two coalitions merge, so $\sum_{i=1}^{n}k_i\leq n - 1$ must hold. Therefore, the total number of oracle accesses is upper bounded by $(n - 1)\left\lceil \log_2 n\right\rceil + 2n-1 \leq n \log_2 n + 3n$.
\end{proofof}



\subsection{Proof of \cref{lem:state}}
\label{appsub:proof_of_lem_state}

\state*

\begin{proofof}{\cref{lem:state}}
    After the initialization in Lines 1 to 4, all of (a), (b), (c) and (d) hold. We will show that after each iteration of the outer while loop, all of (a), (b), (c) and (d) still hold. Thus \cref{lem:state} holds by induction.

    After executing Line 10, $\S$ and $T_{\textup{finalized}}$ are not changed, so (a) and (b) still hold. As $T_i$ is simultaneously changed into $\N\setminus[x]_\S$ for all $i\in[x]_\S$, (c) still holds. According to \cref{lem:auction_gadget}, $\O(\A(\v,\N\setminus[x]_\S)) = \textup{false}$ implies that $\exists j\in \N\setminus[x]_\S$ such that $j\in[x]_{\S^*}$. Thus (d) still holds.

    After executing Line 12, $\S$ and $\mathbf T$ are not changed, so (a), (c) and (d) still hold. According to \cref{lem:auction_gadget}, $\O(\A(\v,\N\setminus[x]_\S)) = \textup{true}$ implies that $[x]_\S\supseteq[x]_{\S^*}$. Using induction hypothesis (a), we can see $[x]_\S=[x]_{\S^*}$. Thus (b) still holds.

    After executing Lines 14 to 18, $\S$ and $T_{\textup{finalized}}$ are not changed, so (a) and (b) still hold. As $T_i$ is simultaneously changed into either $T_\alpha$ or $T_\beta$ for all $i\in[x]_\S$, (c) still holds. According to \cref{lem:auction_gadget}, if $\O(\A(\v,T_\alpha)) = \textup{false}$ then $\exists j\in T_\alpha$ such that $j\in[x]_{\S^*}$. Moreover, using induction hypothesis (d), we can see $j\in[x]_{\S^*}\setminus[x]_\S$. Otherwise, $\O(\A(\v,T_\alpha)) = \textup{true}$. Using induction hypothesis (d) and \cref{lem:auction_gadget}, we can see that $\exists j\in T_\beta$ such that $j\in[x]_{\S^*}\setminus[x]_\S$. In either case, (d) still holds.

    After executing Lines 20 to 22, $T_{\textup{finalized}}$ is not changed, so (b) still holds. Using induction hypothesis (d), we see $[x]_{\S^*}=[y]_{\S^*}$. This shows that (a) still holds. And as $T_i$ is simultaneously changed into $\varnothing$ for all $i\in[x]_\S$, (c) and (d) still hold. This concludes the proof of \cref{lem:state}
\end{proofof}

\subsection{Proof of \cref{lem:termination}}
\label{appsub:proof_of_lem_termination}

\termination*

\begin{proofof}{\cref{lem:termination}}
    We will show that for any $i \in \N$, if $C_i \geq 2\log_2n + 4$ holds for all $j \in [i]_{\S^*}$, then $i \in T_{\textup{finalized}}$ must hold. To prove this, define function $f(T):2^\N\to\mathbb N$ as
    \begin{equation*}
        f(T)=\left\{\begin{array}{ll}
            \lceil\log_2 n\rceil + 1 & (T = \varnothing) \\
            \lceil\log_2 |T|\rceil & (T \ne \varnothing).
        \end{array}\right.
    \end{equation*}
    Let $\S_i = \{[j]_\S\mid j\in [i]_{\S^*}\}$. According to \cref{lem:state} (a), we know that $\S_i$ is a partition of $[i]_{\S^*}$. Moreover, according to \cref{lem:state} (c), for any $S\in\S_i$, $T_x=T_y,\forall x,y\in S$. Therefore, we can unambiguously use $T_S$ to denote $T_x$ for any $x\in S$ and define the potential function $\Phi_i(\mathbf T, \S)$ as
    \begin{equation*}
        \Phi_i(\mathbf T, \S)=\lceil \log_2 n\rceil \cdot |\S_i| + \sum_{S\in\S_i} f(T_S).
    \end{equation*}
    Observe that $\Phi_i(\mathbf T, \S)$ is always a non-negative integer, and after initialization, $\Phi_i(\mathbf T, \S)=(2\lceil \log_2 n\rceil+1)|[i]_{\S^*}|\leq (2 \log_2 n +3)|[i]_{\S^*}|$. To prove \cref{lem:termination}, it suffices to show \textbf{(a)} when $[i]_\S\subset [i]_{\S^*}$, each time $C_j\ (j\in[i]_{\S^*})$ increases, $\Phi_i(\mathbf T, \S)$ decreases by at least $1$, and \textbf{(b)} when $[i]_\S=[i]_{\S^*}$, after $C_j\ (j\in[i]_{\S^*})$ increases again, $i\in T_{\textup{finalized}}$.

    For (a), suppose in Line 7, we get $x \gets \arg\max_{j\in\N}\{v_j\}$ where $x \in [i]_{\S^*}$. If $T_x=\varnothing$, then as $[i]_\S\subset[i]_{\S^*}$, the algorithm must execute Line 10 according to \cref{lem:auction_gadget}. In this case, $f(T_{[x]_\S})$ decreases by at least one. If $T_x\ne\varnothing$, then after Lines 14 to 18, $|T_x|$ decreases by at least $\lfloor |T_x|/2\rfloor$. In this case, $f(T_{[x]_\S})$ also decreases by at least one. Therefore, in either case, $\Phi_i(\mathbf T, \S)$ decreases by at least one after Lines 7 to 18. If the algorithm executes Lines 20 to 22, $[x]_\S$ and $[y]_\S$ merge into one coalition, so $|\S_i|$ decreases by one, and $\Phi_i(\mathbf T, \S)$ decreases by at least $\lceil \log_2 n\rceil+f(T_{[x]_\S})+f([y]_\S)-f(\varnothing)\geq 0$. Thus, after executing Lines 7 to 22, $\Phi_i(\mathbf T, \S)$ decreases by at least one. We then see (a) holds.

    For (b), suppose in Line 7, we get $x \gets \arg\max_{j\in\N}\{v_j\}$ where $x \in [i]_{\S^*}$. As $[i]_\S = [i]_{\S^*}$ already holds, according to \cref{lem:state} (d), $T_x = \varnothing$ must hold. In this case, the algorithm must execute Line 12 according to \cref{lem:auction_gadget}. And $i\in T_\textup{finalized}$ after this line. We then see (b) holds.

    As $C_i = 0,\forall i\in \N$ after initialization, combining (a) and (b), we know when $\sum_{j\in[i]_{\S^*}}C_j\geq (2 \log_2 n +3)|[i]_{\S^*}|+1$, $i\in T_{\textup{finalized}}$ must hold. This proves \cref{lem:termination}.
\end{proofof}

\subsection{Proof of \cref{thm:auction_random_values}}
\label{appsub:proof_of_thm_auction_random_values}

\auctionrandomvalues*

\begin{proofof}{\cref{thm:auction_random_values}}
    According to \cref{lem:state} (b), we know that when $T_{\textup{finalized}} = \N$, $\S = \S^*$ must hold. Therefore, \cref{alg:auction_random_values} always returns the correct answer $\S^*$ after termination. This proves the correctness of \cref{alg:auction_random_values}.
    
    For the sample complexity, \cref{lem:termination} already shows that the expected number of draws to $\V$ is upper bounded by $T_{\textup{ccp}}(n, 2\log_2 n + 4)$, where $T_{\textup{ccp}}(n, m)$ is the expected number of draws needed to collect $m$ sets of coupons when there are $n$ types of coupons. Let $\beta = \frac{2}{\ln 2}$. According to the results of \cite{papanicolaou2020old}, $T_{\textup{ccp}}(n, 2\log_2 n + 4) = (\alpha + o(1))n\ln n$, where $\alpha$ is the unique root of $\alpha-\beta\ln\alpha=\beta-\beta\ln\beta+1$ in $(\beta,+\infty)$. Thus
    \begin{align*}
        T_{\textup{ccp}}(n, 2\log_2 n + 4) &\leq (6.00 + o(1))n\ln n\\
        &\leq (4.16 + o(1))n\log_2 n.
    \end{align*}
    This concludes the proof of \cref{thm:auction_random_values}.
\end{proofof}

\subsection{Proof of \cref{thm:auction_random_special_case}}
\label{appsub:proof_of_thm_auction_random_special_case}

\auctionrandomspecialcase*

\begin{proofof}{\cref{thm:auction_random_special_case}}
    For (a), we will use similar potential analysis to \cref{lem:termination}. Define function $f(T):2^\N\to\mathbb N$ as
    \begin{equation*}
        f(T)=\left\{\begin{array}{ll}
            \lceil\log_2 n\rceil + 1 & (T = \varnothing) \\
            \lceil\log_2 |T|\rceil & (T \ne \varnothing).
        \end{array}\right.
    \end{equation*}
    Now note that when $m = 1$, every agent is in the same coalition. Thus the potential function $\Phi_i(\mathbf T, \S)$ defined in \cref{lem:termination}'s proof is the same for all $i\in\N$. We simplify the notation as $\Phi(\mathbf T, \S)$. That is,
    \begin{equation*}
        \Phi(\mathbf T, \S)=\lceil \log_2 n\rceil \cdot |\S| + \sum_{S\in\S} f(T_S).
    \end{equation*}
    According to \cref{lem:termination}'s proof, $\Phi(\mathbf T, \S)$ is always a non-negative integer, and after initialization, $\Phi(\mathbf T, \S)=(2\lceil \log_2 n\rceil+1)|\S|\leq (2 \log_2 n +3)n$. Moreover, when $\S \ne \S^*$, $\Phi(\mathbf T, \S)$ decreases by at least one after each iteration of the outer while loop. Therefore, when $\S \ne \S^*$, the outer while loop can run for at most $(2 \log_2 n +3)n$ iterations. When $\S = \S^*$, \cref{alg:auction_random_values} terminates in one iteration. Therefore, the sample complexity of \cref{alg:auction_random_values} is bounded by $2n\log_2n+4n$ deterministically.

    For (b), as when $m = n$, every coalition contains only one agent, $\S = \S^*$ after initialization. Therefore, according to \cref{lem:auction_gadget}, for each iteration of the outer while loop, Line 12 must be executed. This means that $T_{\textup{finalized}} = \N$ as soon as we get one item for each agent $i\in\N$ that $i$ values the most. Thus, the expected sample complexity of \cref{alg:auction_random_values} is exactly $T_\textup{ccp}(n,1)$. It is well-known (see e.g. \cite{feller1991introduction}) that $T_\textup{ccp}(n,1) = nH_n$, where $H_n = \sum_{i=1}^{n}\frac{n}{i}$. And as
    \begin{equation*}
        nH_n\leq (1 + o(1))n\ln n\leq (0.70 + o(1))n \log_2 n,
    \end{equation*}
    we conclude that the expected sample complexity of \cref{alg:auction_random_values} is upper bounded by $(0.70 + o(1))n \log_2 n$.
\end{proofof}

\subsection{Auctions Are Not Closed under Product}
\label{appsub:auctions_are_not_closed_under_product}

In the beginning of \cref{subsec:gadget_construction_auction}, we mentioned products of two auctions are no longer auctions. Here, we formalize this statement and provide a simple example to show it.

\begin{proposition}
    \label{prop:auctions_are_not_closed_under_product}
    There exist a set of agents $\N$ and two auctions $\A_\alpha,\A_\beta$ among $\N$ such that $\A_\alpha\times \A_\beta$ is not an auction. 
\end{proposition}

\begin{proofof}{\cref{prop:auctions_are_not_closed_under_product}}
    We show this statement with a concrete example. Let $\N=\{1,2\}$ be the set of bidders and $\A_\alpha = (\v_\alpha, \r_\alpha),\A_\beta = (\v_\beta, \r_\beta)$ be two second price auctions with personalized reserves among $\N$. Let $\v_\alpha = (1, 0), \v_\beta = (0, 1)$ and $\r_\alpha = \r_\beta = (0, 0)$. Assume bidders $1$ and $2$ are not in the same coalition.

    Consider the product of the auctions $\A_\alpha\times \A_\beta$. If $\A_\alpha\times \A_\beta$ is also an auction, then only one of the players' final utility can be positive. However, if player $1$ bids $(\frac{1}{2},0)$ and player $2$ bids $(0,\frac{1}{2})$ in $\A_\alpha\times \A_\beta$, both player end up with $\frac{1}{2}$ utility in $\A_\alpha\times \A_\beta$. This shows that, $\A_\alpha\times \A_\beta$ is not an auction.
\end{proofof}

\section{CSL with Graphical Games}
\label{app:csl_with_graphical_games}

In this section, we discuss how to extend \cref{alg:normal_form_games} to solve the CSL problem with graphical games. We will refer to this problem as the GraphicalCSL problem below.

Recall that a graphical game is represented by a graph $G$, where each vertex denotes a player. There is an edge between a pair of vertices $x$ and $y$ if and only if their utilities are dependent on each other's strategy. To limit the size of representation of a graphical game, a common way is to limit the maximum vertex degree $d$ in $G$. When $d = n$, any normal form game can be represented by a graphical game. In this case, \cref{alg:normal_form_games} can be directly applied to solve the GraphicalCSL problem. We show in this section that a slight variant of \cref{alg:normal_form_games} can also be used to solve the GraphicalCSL problem when $d = 1$. Note that $d = 1$ is the most restrictive case, as it means that the utilities of all players can only depend on the strategy of one other player. Additionally, graphical games with $d = 1$ are also a subset of polymatrix games, so this algorithm can also be used to solve the CSL problem with Polymatrix Games.

To start with, we still want to use the product of normal form gadgets (as defined in \cref{def:normal_form_gadget,def:product_of_normal_form_games}) as our primary building block. However, as we additionally require the games we use to be graphical games (with $d = 1$), we need to make sure that the products of normal form gadgets we use are also graphical games. We establish a sufficient condition for this in the following \cref{lem:graphical_game_gadget}.

\begin{lemma}
    \label{lem:graphical_game_gadget}
    Let $k\in\mathbb N^{+}$ and $\x,\y\in\N^{k}$ be two arrays of agents of length $k$. If all $2k$ agents in $\x$ and $\y$ are distinct, then $\prod_{\theta=1}^{k}\NF(x_\theta,y_\theta)$ is a graphical game with $d = 1$.
\end{lemma}

\begin{proofof}{\cref{lem:graphical_game_gadget}}
    In a normal form game gadget $\NF(x,y)$, the utilities of $x$ and $y$ are dependent on each other's strategy, while for any $i\in\N,i\ne x,y$, the utility of $i$ is independent of other agents' strategy. As all $2k$ agents in $\x$ and $\y$ are distinct, in the product game $\prod_{\theta=1}^{k}\NF(x_\theta,y_\theta)$, the utilities of $x_\theta$ and $y_\theta$ are only dependent on each other's strategy for all $\theta\in\{1,\dots,k\}$. The lemma then follows.
\end{proofof}

For $\x,\y\in\N^k$, if we view $(\x,\y)$ as an undirected graph with $n$ vertices and $k$ edges $(x_\theta,y_\theta)$, then \cref{lem:graphical_game_gadget} essentially shows that as long as $(\x,\y)$ forms a graph matching, $\prod_{\theta=1}^{k}\NF(x_\theta,y_\theta)$ is a graphical game with $d = 1$. To proceed and show our algorithm, we invoke the following well-known fact about graph matchings.

\begin{lemma}
    \label{lem:graph_decomposition}
    Let $n$ be a positive integer, and let $K_n$ denote the complete graph with $n$ vertices. Then
    \begin{enumerate}[(a)]
        \item If $n$ is even, there are $n - 1$ matchings $\{M_1,\dots,M_{n-1}\}$ where $M_\theta = \{(x_{\theta,1},y_{\theta,1}),\dots,(x_{\theta,{\frac{n}{2}}},y_{\theta,{\frac{n}{2}}})\}$, such that each edge in $K_n$ is in exactly one matching.
        \item If $n$ is odd, there are $n$ matchings $\{M_1,\dots,M_{n}\}$ where $M_\theta = \{(x_{\theta,1},y_{\theta,1}),\dots,(x_{\theta,\frac{n-1}{2}},y_{\theta,\frac{n-1}{2}})\}$, such that each edge in $K_n$ is in exactly one matching.
    \end{enumerate}
\end{lemma}

\cref{lem:graph_decomposition} is well-known in the graph one-factorization and round-robin tournament scheduling literature. For a proof of \cref{lem:graph_decomposition}, we refer to a recent tutorial \cite{ribeiro2023tutorial}. With \cref{lem:graphical_game_gadget,lem:graph_decomposition}, we are now ready to present our algorithm below.

\begin{algorithm}[htbp]
    \caption{IG with Graphical Games (GraphicalIG)}
    \label{alg:graphical_games}

    \textbf{Input}: The number of agents $n$ and an observation oracle $\O$

    \textbf{Output}: The coalition structure $\S$ of the agents

    \begin{algorithmic}[1] 
        \STATE Let $\S\gets\{\{1\},\{2\},\dots,\{n\}\}$.
        \STATE Let $\M$ be the set of matchings as in \cref{lem:graph_decomposition}.
        \FOR {$M\in\M$}
            \WHILE {true}
                \STATE Let $T\gets\{(x,y)\in M\mid [x]_\S \ne [y]_\S\}$.
                \IF {$\O(\prod_{(x,y)\in T}\NF(x,y)) = \textup{true}$}
                    \STATE \textbf{break}.
                \ENDIF
                \WHILE {$|T| > 1$}
                    \STATE Partition $T$ into $T_\alpha,T_\beta$ where $\left||T_\alpha|-|T_\beta|\right|\leq 1$.
                    \IF {$\O(\prod_{(x,y)\in T_\alpha}\NF(x,y)) = \textup{false}$}
                        \STATE Let $T\gets T_\alpha$.
                    \ELSE
                        \STATE Let $T\gets T_\beta$.
                    \ENDIF
                \ENDWHILE
                \STATE Let $(x,y)\gets \textup{the only element in $T$}$.
                \STATE Merge $[x]_\S$ and $[y]_\S$ in $\S$.
            \ENDWHILE
        \ENDFOR
        \STATE \textbf{return} $\S$.
    \end{algorithmic}
\end{algorithm}

GraphicalIG follows the same high-level idea, i.e., iteratively merging coalitions found with binary search, as \cref{alg:normal_form_games}. In each iteration of the outer for loop (Lines 4 to 15), the algorithm picks a matching $M$ and only uses products of normal form gadgets that correspond to edges in $M$. This ensures that the products of normal form gadgets used in each iteration are graphical games with $d = 1$ according to \cref{lem:graphical_game_gadget}. The algorithm then proceeds with an infinite while loop (Line 4). In each iteration of this loop (Lines 5 to 15), the algorithm tries to identify a pair $(x,y)\in M$ such that $[x]_{\S^*} = [y]_{\S^*}$, but currently $[x]_\S \ne [y]_\S$. It first picks all pairs $(x,y)\in M$ such that $[x]_\S \ne [y]_\S$ as $T$ (Line 5), and then it checks whether any pair of agents in $T$ is in the same coalition (Line 6). If there are no such pairs, it breaks the infinite loop (Line 7). Otherwise, it uses a binary search process similar to \cref{alg:normal_form_games} to find one of such pairs $(x, y)$ (Lines 8 to 14). After finding such a pair, it merges $[x]_\S$ and $[y]_\S$ in $\S$ (Line 15). The outer for loop then repeats this process for all matchings in $\M$. As edges in $\M$ covers all possible pairs of agents, the algorithm will eventually exhaust all pairs and find the correct coalition structure $\S^*$.

We show the following guarantees for GraphicalIG.

\begin{theorem}
    \label{thm:graphical_games}
    GraphicalIG solves GraphicalCSL with a sample complexity upper bounded by $n \log_2 n + 3n$. 
\end{theorem}

\begin{proofof}{\cref{thm:graphical_games}}
    For the correctness of GraphicalIG, consider one iteration of Lines 5 to 15. Let $M$ be the matching used in this iteration. First of all, as $M$ is a matching and the algorithm only uses products of normal form gadgets that correspond to edges in $M$, according to \cref{lem:graphical_game_gadget}, the algorithm only uses graphical games with $d = 1$. If $\exists (x,y)\in M$ such that $[x]_{\S^*} = [y]_{\S^*}$ and $[x]_{\S}\ne[y]_{\S}$, then such pairs will be included in $T$ (Line 5). Moreover, as the algorithm uses a binary search process similar to \cref{alg:normal_form_games} to find one of such pairs $(x, y)$ (Lines 8 to 14), it will eventually find such a pair and merge $[x]_\S$ and $[y]_\S$ in $\S$ (Line 15). Otherwise, if no such pairs exist, the algorithm will break the infinite loop (Line 7) according to \cref{lem:product_of_normal_form_game_gadget}. Therefore, we can see that (i) if $[x]_{\S^*} \ne [y]_{\S^*}$, $(x,y)$ never gets to be merged in Line 14, (ii) if $[x]_{\S^*} \ne [y]_{\S^*}$ and $(x,y)\in M$, then the execution of Lines 4 to 15 ensures that $[x]_\S = [y]_\S$ afterward. Combining (i) and (ii), and as matchings in $\M$ cover all possible pairs of agents, we see that GraphicalIG will find the correct coalition structure $\S^*$.

    For the complexity of GraphicalIG, consider the following two cases. If $\O(\prod_{(x,y)\in T}\NF(x,y)) = \textup{true}$ on Line 6, the infinite loop breaks. So this happens once for each matching $M\in\M$, resulting in $|\M|$ oracle accesses. If $\O(\prod_{(x,y)\in T}\NF(x,y)) = \textup{false}$ on Line 6, then necessarily, we will find a pair of agents $(x,y)$ and merge their coalitions in $\S$, resulting in a decrease in the number of coalitions in $\S$ by one. Therefore, this happens at most $n - 1$ times. Each time this happens, we do a binary search, and the number of oracle accesses is upper bounded by $1 + \lceil \log_2 n \rceil \leq \log_2 n + 2$. Combining the two cases, we see that the number of oracle accesses is upper bounded by $|\M| + (n - 1)(\log_2 n + 2)$. According to \cref{lem:graph_decomposition}, $|\M| \leq n$. Therefore, the number of oracle accesses is upper bounded by $n \log_2 n + 3n$.
\end{proofof}

\cref{thm:graphical_games} shows that using graphical games with the most restrictive contraint $d = 1$, we can still solve CSL with GraphicalIG within a sample complexity of $n \log_2 n + 3n$.

\section{CSL with Designed Auctions}
\label{app:csl_with_designed_auctions}

In this section, we consider a simplified the setting of AuctionCSL where we can specify the agent valuations $\{v_i\}$. As we will see, this allows an algorithm very similar to IG with the same sample optimal complexity $n\log_2 n + 3n$.


Concretely, suppose there are $n$ types of items. The $i$-th item is only valuable to the $i$-th agent, i.e., the valuation vector $\v^{(i)}$ is defined as $v^{(i)}_j=\ind{j = i}$.
We assume in this subsection that we have the freedom to choose any one of the $n$ items to auction each time. 
In practice, we not only care about the sample complexity (which is the total number of used items) of the algorithm but also the maximum number of items of each type we need to use. This is because the number of items of each type is often limited in practice. With that in mind, we proceed with the following algorithm.




\begin{algorithm}[htbp]
    \caption{IG with Designed Auctions (DAIG)}
    \label{alg:auction_partial_control}

    \textbf{Input}: The number of agents $n$ and an observation oracle $\O$

    \textbf{Output}: A coalition structure $\S$ of the agents

    \begin{algorithmic}[1] 
        \STATE Let $\S\gets\{\{1\},\{2\},\dots,\{n\}\}$.
        \FOR {$i\in\N$}
            \STATE Let $x \gets i$ and let $\v \gets \v^{(i)}$.
            \WHILE {$\O(\A(\v,\N\setminus[x]_\S)) = \textup{false}$}
                \STATE Let $T \gets \N\setminus[x]_\S$.
                \WHILE {$|T| > 1$}
                    \STATE Partition $T$ into $T_\alpha,T_\beta$ where $\left||T_\alpha|-|T_\beta|\right|\leq 1$.
                    \STATE Let $\v \gets \v^{(x)}$.
                    \IF {$\O(\A(\v,T_\alpha)) = \textup{false}$}
                        \STATE Let $T\gets T_\alpha$.
                    \ELSE
                        \STATE Let $T\gets T_\beta$.
                    \ENDIF
                \ENDWHILE
                \STATE Let $y\gets \textup{the only element in $T$}$.
                \STATE Merge $[x]_\S$ and $[y]_\S$ in $\S$.
                \STATE Let $x \gets y$ and let $\v \gets \v^{(x)}$.
            \ENDWHILE
        \ENDFOR
        \STATE \textbf{return} $\S$.
    \end{algorithmic}
\end{algorithm}

DAIG also follows the same high-level idea as IG (\cref{alg:normal_form_games}). In each iteration of the outer for loop (Lines 3 to 15), we also try to find all players in $i$'s coalition $[i]_{\S^*}$ with the help of auction gadgets and \cref{lem:auction_gadget}. The main difference here is that for each auction we run, we will consume an item. Therefore, we do not always use type $i$ items in the $i$-th iteration of the outer for loop. Instead, we let $x$ be $i$ (Line 3) and try to find another agent $y$ in the same coalition with $x$ by using type $x$ items (Lines 4 to 13). If such $y$ is found, we update the coalition structure (Line 14) and let $x$ be $y$ (Line 15) to continue the search. In this way, we start to use type $y$ items in the search, which results in a more balanced use of items of different types.

\begin{restatable}{theorem}{auctionpartialcontrol}
    \label{thm:auction_partial_control}
    DAIG solves the simplified AuctionCSL with a sample complexity upper bounded by $n \log_2 n + 3n$. Moreover, $\forall i\in\N$, it uses at most $\log_2 n + 3$ items of the $i$-th type.
\end{restatable}

\begin{proofof}{\cref{thm:auction_partial_control}}
    The correctness proof of DAIG is essentially the same as the correctness proof of IG (\cref{alg:normal_form_games}). Let the ground truth coalition structure of the agents be $\S^*$. We can still show that after the $i$-th iteration of the outer for loop, $[j]_\S=[j]_{\S^*},\forall j\leq i$ and $[j]_\S\subseteq[j]_{\S^*},\forall j > i$ by induction. The only difference is that we use \cref{lem:auction_gadget} instead of \cref{lem:product_of_normal_form_game_gadget} to check whether there is an agent in $T$ that is in the same coalition with $i$. The correctness proof of DAIG then follows from the correctness proof of IG (\cref{thm:normal_form_games}).

    Next, we prove the sample complexity of DAIG. We will directly show that DAIG uses at most $\log_2 n + 3$ type $i$ items for $i\in\N$. The sample complexity upper bound is then implied by this. To see this, consider the following two cases. \textbf{(1).} If $i$ is the first element in $[i]_{\S^*}$, then the while loop body (Lines 5 to 15) will be executed until $[i]_{\S}=[i]_{\S^*}$. As we only use type $x$ items in the loop body, and $x$ is set to $y$ (which is the newly indentified member in $[x]_\S$) in Line 15, items of type $j$ are used no more than $\lceil \log_2 n \rceil + 1\leq \log_2 n + 2$ times for each $j\in[i]_{\S^*}$. \textbf{(2).} If $i$ is not the first element in $[i]_{\S^*}$. Then $[i]_{\S}=[i]_{\S^*}$ already holds before the $i$-th iteration in the out for loop. So the while loop body (Lines 5 to 15) will not be executed according to \cref{lem:auction_gadget}. In this case, only $1$ type $i$ item is used. Combining (1) and (2), we see that at most $\log_2 n + 3$ items of each type are used by DAIG. This concludes the proof of \cref{thm:auction_partial_control}.
\end{proofof}

\cref{thm:auction_partial_control} shows that  we can solve the simplified AuctionCSL with the same sample complexity as in the normal form game setting. Moreover, the number of items used of each type is also bounded to $\log_2n + o(1)$.

\end{document}